\documentclass[pre,aps,showkeys,showpacs,superscriptaddress,twocolumn,%
amsmath,amssymb]{revtex4} \usepackage{graphicx}

\newcommand{\df}{\Delta F}
\newcommand{\dfhat}{\widehat{\df}}
\newcommand{\dfhatb}{\dfhat_{01}}
\newcommand{\dfhatf}{\dfhat_0}
\newcommand{\dfhatr}{\dfhat_1}
\newcommand{\dfhatB}{\dfhat_{\text{B}}}
\newcommand{\dfhatftrad}{\dfhat^{\text{trad}}_0}
\newcommand{\dfhatrtrad}{\dfhat^{\text{trad}}_1}
\newcommand{\dhtilde}{\widetilde{\Delta H}}
\newcommand{\x}{{ x}}
\newcommand{\y}{{ y}}
\newcommand{\phir}{{\phi}^{-1}}
\newcommand{\M}{\mathcal{M}}
\newcommand{\e}{\mathrm{e}}
\newcommand{\widebar}[1]{\overline{#1}}
\newcommand{\rhotildef}{\tilde{\rho}_0}
\newcommand{\Gammatildef}{\tilde{\Gamma}_0}
\newcommand{\lnlikeli}{\ln\mathcal{L}}
\newcommand{\gl}[1]{\eqref{#1}} 
\newcommand{\rr}{{\bf r}}
\newcommand{\RR}{{\bf R}}
\newcommand{\MSE}{\operatorname{mse}}
\newcommand{\muex}{\mu^{\text{ex}}}
\newcommand{\muhat}{\widehat{\muex}}
\newcommand{\muhatb}{\muhat_{01}}
\newcommand{\muhatf}{\muhat_0}
\newcommand{\muhatr}{\muhat_1}
\newcommand{\N}{{N_p}}
\newcommand{\Npl}{{N_p+1}}
\newcommand{\Rb}{R_{\text{box}}}
\newcommand{\aol}{A_{\text{ol}}}
\newcommand{\aolhat}{\hat{A}_{\text{ol}}}
\newcommand{\aolhatv}{\aolhat^{\scriptscriptstyle(I\!I)}}
\newcommand{\la}{\left\langle}
\newcommand{\ra}{\right\rangle}
\newcommand{\id}{\text{\it id.}}
\newcommand{\rdf}{\text{\it rdf}}
 
\begin{document}

\title{Using bijective maps to improve free energy estimates}
\author{A.M.\,Hahn}
\affiliation{Institut f\"ur Physik, Carl von Ossietzky Universit\"at,
 26111 Oldenburg, Germany}
\author{H.\,Then}
\affiliation{Institut f\"ur Physik, Carl von Ossietzky Universit\"at,
 26111 Oldenburg, Germany}

\begin{abstract}
 We derive a fluctuation theorem for generalized work distributions,
 related to bijective mappings of the phase spaces of two physical
 systems, and use it to derive a two-sided constraint maximum
 likelihood estimator of their free energy difference which uses
 samples from the equilibrium configurations of both systems. As an
 application, we evaluate the chemical potential of a dense
 Lennard-Jones fluid and study the construction and performance of
 suitable maps.
\end{abstract}

\pacs{05.40.-a, 05.70.Ln}
\keywords{fluctuation theorem, nonequilibrium thermodynamics}

\maketitle

\section{Introduction}

Extracting free energy differences from a suitable set of computer
simulation data is an active field of research and of interest for
e.g. drug design \cite{Reddy2001} or nonperturbative quantum
chromodynamics \cite{Schafer1998}. Concerning estimators for the free
energy difference, an extensive literature can be found. Probably the
most elementary estimator is the traditional free energy perturbation
\cite{Zwanzig1954}, which is briefly introduced in the following.
Assume we have given two systems, arbitrarily labeled as system $0$
and system $1$, that are characterized by Hamiltonians $H_0(\x)$ and
$H_1(\x)$, respectively, depending on the point $\x$ in phase space.
Further, let $\rho_i(\x)$ denote the thermal equilibrium phase space
density of system $i$,
\begin{align} \label{rhodef}
 \rho_i(\x) = \frac{\e^{-\beta H_i(\x)}}{Z_i}, \quad i=0,1,
\end{align}
where $Z_i=\int \e^{-\beta H_i(\x)} d\x$ denotes the partition
function and $\beta=\frac{1}{kT}$ the inverse temperature. We are
interested in the Helmholtz free energy difference $\df$ of the
systems, defined as $\df = -\frac{1}{\beta} \ln \frac{Z_1}{Z_0}$.
Traditional free energy perturbation \cite{Zwanzig1954} originates
from the equality
\begin{align} \label{rhodrho}
 \frac{\rho_0(\x)}{\rho_1(\x)} = \e^{\beta (\Delta H(\x)- \df)},
\end{align}
with $\Delta H(\x):= H_1(\x)-H_0(\x)$. The latter quantity may be
interpreted as the work performed during an infinitely fast switching
process transforming system $0$ to system $1$, with initial
configuration $\x$ \cite{Jarzynski1997}. A direct consequence of
Eq.~\gl{rhodrho} is the perturbation identity
\begin{align} \label{pert0} \e^{-\beta \df}=\int \e^{-\beta \Delta
 H(\x)} \rho_0(\x) d\x,
\end{align}
which is frequently used to obtain an estimate of $\df$ in drawing a
sample $\{\x_1,\dots,\x_N\}$ from $\rho_0(\x)$ (e.g. by Monte Carlo
simulations) and evaluating the estimator
\begin{align} \label{dfhatftrad} \dfhatftrad = -\frac{1}{\beta} \ln
 \widebar{\e^{-\beta\Delta H(\x)}}.
\end{align}
The overbar denotes a sample average [ i.e.
$\widebar{f(\x)}=\frac{1}{N}\sum_{k=1}^{N}f(\x_k)$ where $f$ stands
for an arbitrary function]. As can be seen by comparison with
Eq.~\gl{rhodrho}, the integrand appearing in Eq.~\gl{pert0} is
proportional to $\rho_1$, and thus the main contributions to an
accurate estimate of $\df$ with Eq.~\gl{dfhatftrad} come from
realizations $\x$ (drawn from $\rho_0$) that are typical for the
density $\rho_1$. This means that the performance of such an estimate
depends strongly on the degree of overlap of $\rho_0$ with $\rho_1$.
If the overlap is small, the traditional free energy perturbation is
plagued with a slow convergence and a large bias. This can be
overcome by using methods that bridge the gap between the densities
$\rho_0$ and $\rho_1$, for instance the thermodynamic integration.
Since thermodynamic integration samples a sequence of many equilibrium
distributions, it soon becomes computationally expensive. Another
method is umbrella sampling \cite{Torrie1977} which distorts the
original distribution in order to sample regions that are important
for the average. Because of the distortion, the latter method is in
general restricted to answer only one given question, e.g. the value
of the free energy difference, but fails to give further answers.
This is of particular concern, if in addition the values of some other
thermodynamic variables are sought, for example pressure or internal
energy. There are dynamical methods \cite{Frenkel2002} that make use
of the Jarzynski work theorem \cite{Jarzynski1997}. They allow to base
the estimator on work values of fast, finite time, non-equilibrium
processes connecting system $0$ with system $1$. However, the dynamic
simulation of the trajectories is typically very expensive.

Six years ago, the targeted free energy perturbation method
\cite{Jarzynski2002} was introduced; a promising method which is based
on mapping equilibrium distributions close to each other in order to
overcome the problem of insufficient overlap, without the need to draw
from biased distributions. However, this method is hardly used in the
literature. An obstacle might be that there is no general description
of how to construct a suitable map. A recent improvement is the
escorted free energy simulation \cite{Vaikuntanathan2008} which is a
dynamical generalization of the targeted free energy perturbation.

Any free energy difference refers to two equilibrium ensembles. The
above mentioned methods draw only from one of the two 
ensembles and propagate the system in direction of the other.
Insofar, they are ``one-sided'' methods. However, it is of advantage
to draw from both equilibrium distributions and combine the obtained
``two-sided'' information. Optimizing the elementary ``two-sided''
estimator for free energy differences results in the acceptance ratio
method \cite{Bennett1976,Crooks2000,Shirts2003}. The next step of
improvement is to implement a ``two-sided'' targeted free energy
method that optimally employs the information of drawings from both
equilibrium distributions. Our aim is to combine the advantages of
the acceptance ratio method with the advantages of the targeted free
energy perturbation.

The central result of this paper is a fluctuation theorem for the
distributions of generalized work values that is derived and presented
in section \ref{sec-ft}. From this fluctuation theorem, the desired
optimal two-sided targeted free energy estimator follows in section
\ref{sec-est}. In section \ref{sec-rms}, appropriate measures are
introduced which relate the overlap of $\rhotildef$ with $\rho_1$ to
the mean square errors of the one- and two-sided free energy
estimators. In section \ref{sec-con}, a convergence criterion for the
two-sided estimator is proposed. From section \ref{sec-num} on,
numerics plays an important part. In particular, section
\ref{sec-cavity} deals with explicit numerical applications. Based on
the two-sided targeted free energy estimator, in section \ref{sec-mu},
an estimator for the chemical potential of a high-density homogeneous
fluid is established and applied to a dense Lennard-Jones fluid.
Finally, the construction and performance of suitable maps is studied.

In order to get some notation straight, we start by recalling the
targeted free energy perturbation method.

\section{Targeted free energy perturbation}

Let $\Gamma_0$ and $\Gamma_1$ denote the phase spaces of the systems
$0$ and $1$, respectively. We require that $\Gamma_i$ contains only
those points $\x$ for which $\rho_i(\x)$ is non-zero.

Mapping the phase space points of system $0$, $\x\to\phi(\x)$, such
that the mapped phase space $\Gammatildef=\phi(\Gamma_0)$ coincides
with the phase space $\Gamma_1$ and such that the mapped distribution
$\rhotildef$ overlaps better with the canonical distribution $\rho_1$
results in the {\em targeted free energy perturbation}
\cite{Jarzynski2002} where the samples are drawn {\em effectively}
from $\rhotildef$, instead.

Following the idea of Jarzynski \cite{Jarzynski2002}, we introduce
such a phase space map. If $\Gamma_0$ and $\Gamma_1$ are diffeomorph,
there exists a bijective and differentiable map $\M$ from $\Gamma_0$
to $\Gamma_1$,
\begin{align}
 \M: \Gamma_0 \to \Gamma_1, \qquad \M: \x \to \phi(\x),
\end{align}
where the absolute value of the Jacobian is
\begin{align} \label{kdef} K(\x)= | \Big| \frac{\partial
 \phi}{\partial \x} \Big| |.
\end{align}
The inverse map reads
\begin{align}
 \M^{-1}: \Gamma_1 \to \Gamma_0, \qquad \M^{-1}: \x \to \phir(\x).
\end{align}

According to $\mathcal M$, the phase space density $\rho_0$ is mapped
to the density $\rhotildef$,
\begin{align} \label{etadef1} \rhotildef(\y) = \int_{\Gamma_0}
 \delta(\y-\phi(\x)) \rho_0(\x) d\x,
\end{align}
which can be written as
\begin{align} \label{etadef2}
 \rhotildef(\phi(\x)) = \frac{\rho_0(\x)}{K(\x)}
\end{align}
or
\begin{align} \label{etadef4}
 \int_{\phi(\Gamma)} \rhotildef(\y) d\y = \int_{\Gamma}\rho_0(\x) d\x,
 \quad \forall \Gamma \subset \Gamma_0.
\end{align}

In analogy to Eq.~\gl{rhodrho}, the targeted free energy perturbation
is based on the identity
\begin{align} \label{etadrho}
 \frac{\rhotildef(\phi(\x))}{\rho_1(\phi(\x))} = \e^{\beta(
 \dhtilde(\x) - \df)} \quad \forall \x \in \Gamma_0,
\end{align}
which follows from the densities \gl{rhodef} and \gl{etadef2} with
$\dhtilde$ being defined by
\begin{align} \label{dhtildedef}
 \dhtilde(\x) := H_1(\phi(\x)) - H_0(\x) - \frac{1}{\beta} \ln K(\x).
\end{align}
Multiplying Eq.~\gl{etadrho} by $\e^{-\beta\dhtilde(\x)}
\rho_1(\phi(\x)) K(\x)$ and integrating over $\Gamma_0$ yields the
targeted free energy perturbation formula,
\begin{align} \label{targid0}
 \e^{-\beta \df}= \int_{\Gamma_0} \e^{-\beta \dhtilde(\x)} \rho_0(\x)
 d\x .
\end{align}
An alternative derivation is given in \cite{Jarzynski2002}. The
traditional free energy perturbation formula \gl{pert0} can be viewed
as a special case of Eq.~\gl{targid0}. The latter reduces to the
former if $\M$ is chosen to be the identity map, $\phi(\x)=\x$. [This
requires that $\Gamma_1=\Gamma_0$ holds.]

Now an obvious estimator for $\df$, given a sample $\{\x_k\}$ drawn
from $\rho_0(\x)$, is
\begin{align} \label{dfhat0targ} \dfhatf = -\frac{1}{\beta} \ln
 \widebar{\e^{-\beta\dhtilde(\x)}},
\end{align}
which we refer to as the targeted {\em forward} estimator for $\df$.
The convergence problem of the traditional forward estimator,
Eq.~\gl{dfhatftrad}, in the case of insufficient overlap of $\rho_0$
with $\rho_1$ is overcome in the targeted approach by choosing a
suitable map $\M$ for which the image $\rhotildef$ of $\rho_0$
overlaps better with $\rho_1$. Indeed, suppose for the moment that the
map is chosen to be ideal, namely such that $\rhotildef(\x)$ coincides
with $\rho_1(\x)$. Then, as a consequence of Eq.~\gl{etadrho}, the
quantity $\dhtilde(\x)$ is constant and equals $\df$, and the
convergence of the targeted estimator \gl{dfhat0targ} is immediate.
Although the construction of such an ideal map is impossible in
general, the goal of approaching an ideal map guides the design of
suitably good maps.

To complement the {\em one-sided} targeted estimator, a second
perturbation formula in the ``reverse'' direction is derived from
Eq.~\gl{etadrho},
\begin{align} \label{targid1}
 \e^{+\beta \df} = \int_{\Gamma_1} \e^{+\beta \dhtilde(\phir(\y))}
 \rho_1(\y) d\y,
\end{align}
leading to the definition of the targeted {\em reverse} estimator
$\dfhatr$ of $\df$,
\begin{align} \label{dfhat1targ} \dfhatr = +\frac{1}{\beta} \ln
 \widebar{\e^{+\beta\dhtilde(\phir(\y)}}.
\end{align}
The index $1$ indicates that the set $\{\y_k\}$ is drawn from
$\rho_1$. Using the identity map $\phi(\x)=\x$ in Eq.~\gl{dfhat1targ}
gives the traditional reverse estimator, which is valid if
$\Gamma_0=\Gamma_1$ holds.

It will prove to be beneficial to switch from phase space densities to
one-dimensional densities which describe the value distributions of
$\dhtilde(\x)$ and $\dhtilde(\phir(\y))$, cf. Eqs.~\gl{targid0} and
\gl{targid1}. This is done next and results in the fluctuation theorem
for generalized work distributions.

\section{Fluctuation theorem for generalized work distributions}
\label{sec-ft}

We call $\dhtilde(\x)$, $\x\in\Gamma_0$, function of the {\em
 generalized work} in {\em forward} direction and
$\dhtilde(\phir(\y))$, $\y\in\Gamma_1$, function of the {\em
 generalized work} in {\em reverse} direction, having in mind that
these quantities are the functions of the actual physical work for
special choices of the map $\M$ \cite{Schoell-Paschinger2006}.

The probability density $p(W|0;\M)$ for the outcome of a specific
value $W$ of the generalized work in forward direction subject to the
map $\mathcal M$ when sampled from $\rho_0$ is given by
\begin{align} \label{pfdef} p(W|0;\M) = \int_{\Gamma_0} \delta
 (W-\dhtilde(\x)) \rho_0(\x) d\x.
\end{align}
Conversely, the probability density $p(W|1;\M)$ for the observation of
a specific value $W$ of the generalized work in reverse direction when
sampled from $\rho_1$ reads
\begin{align} \label{prdef}
 p(W|1;\M) = \int_{\Gamma_1} \delta (W-\dhtilde(\phir(\y)))
 \rho_1(\y) d\y.
\end{align}
Relating the forward and reverse ``work'' probability densities to
each other results in the fluctuation theorem
\begin{align} \label{crooks} \frac{p(W|0;\M)}{p(W|1;\M)} = \e^{\beta(
 W -\df)}.
\end{align}
This identity provides the main basis for our further results. It is
established by multiplying Eq.~\gl{etadrho} with
$\delta(W-\dhtilde(\x))\rho_1(\phi(\x))$ and integrating with respect
to $\phi(\x)$. The left hand side yields
\begin{multline}
 \int_{\phi(\Gamma_0)} \delta(W-\dhtilde(\x)) \rhotildef(\phi(\x))
 d\phi(\x)
 \\
 =\int_{\Gamma_0} \delta(W-\dhtilde(\x)) \rho_0(\x) d\x = p(W|0;\M),
\end{multline}
and the right hand side gives
\begin{multline}
 \int_{\phi(\Gamma_0)} \e^{\beta(\dhtilde(\x) - \df)}
 \delta(W-\dhtilde(\x)) \rho_1(\phi(\x)) d\phi(\x)
 \\
 = \e^{\beta (W - \df)} \int_{\Gamma_1} \delta(W-\dhtilde(\phir(\y)))
 \rho_1(\y) d\y
 \\
 = \e^{\beta (W - \df)} p(W|1;\M).
\end{multline}

It is worth to emphasize that the fluctuation theorem \gl{crooks} is
an exact identity for {\em any} differentiable, bijective map $\M$
from $\Gamma_0$ to $\Gamma_1$. Especially, it covers known fluctuation
theorems \cite{Crooks1999,Jarzynski2000,Evans2003,Cuendet2006} related
to the {\em physical} work applied to a system that is driven
externally and evolves in time according to some deterministic
equations of motion, e.g. those of Hamiltonian dynamics,
Nos\'{e}-Hoover dynamics or Gaussian isokinetic dynamics
\cite{Schoell-Paschinger2006}.

As an example, consider the time-reversible adiabatic evolution of a
conservative system with Hamiltonian $H_\lambda(\x)$, depending on an
externally controlled parameter $\lambda$ (e.g. the strength of an
external field). Let $\x(t)=\phi(\x_0,t;\lambda(\cdot))$ with
$\x(0)=\x_0$ be the flow of the Hamiltonian system which is a
functional of the parameter $\lambda(t)$ that is varied from
$\lambda(0)=0$ to $\lambda(\tau)=1$ according to some prescribed
protocol that constitutes the forward process. The Hamiltonian flow
can be used to define a map,
$\M:\x\to\phi(\x):=\phi(\x,\tau;\lambda(\cdot))$. Since the evolution
is adiabatic and Hamiltonian, no heat is exchanged, $Q=0$, and the
Jacobian is identical to one, $|\frac{\partial \phi}{\partial \x}|=1$.
Consequently, the generalized work in forward direction reduces to the
physical work applied to the system,
$\dhtilde(\x)=H_1(\phi(\x))-H_0(\x)=W$. For each forward path
$\{\x(t),\lambda(t);\ 0\le t\le\tau\}$ we have a reverse path
$\{\x^T(\tau-t),\lambda^T(\tau-t);\ 0\le t\le\tau\}$, where the
superscript $T$ indicates that quantities that are odd under time
reversal (such as momenta) have changed their sign. The generalized
work in reverse direction reduces to the physical work done {\em by}
the system,
$-W=-(H_0(\phir(y)^T)-H_1(y^T))=-H_0(\phir(y))+H_1(y)=\dhtilde(\phir(y))$.
Starting the forward process with an initial canonical distribution,
$\rho_0(\x)$, some probability distribution for the physical work in
forward direction follows, $p(W|0;\M)=:p^F(W)$. Starting the reverse
process with an initial canonical distribution, $\rho_1(\y)$, some
probability distribution for the physical work in reverse direction
follows, $p(W|1;\M)=:p^R(-W)$. The distributions $p^F(W)$ and
$p^R(-W)$ are related to each other by the identity \gl{crooks} which
coincides with the fluctuation theorem of Crooks \cite{Crooks1999}.

From the fluctuation theorem \gl{crooks} some important inequalities
follow that are valid for any map $\M$. First of all we state that
the targeted free energy perturbation formulas \gl{targid0} and
\gl{targid1} can be regarded as a simple consequence of the
fluctuation theorem \gl{crooks} and can be rewritten in terms of the
generalized work distributions, $\e^{-\beta\df}=\la \e^{-\beta
 W}\ra_0$ and $\e^{+\beta\df}=\la \e^{+\beta W}\ra_1$, where the
angular brackets with subscript $i$ denote an ensemble average with
respect to the density $p(W|i;\M)$, $i=0,1$. The monotonicity and convexity of the
exponential function appearing in the above averages allows the
application of Jensen's inequality, $\la \e^{\mp\beta W} \ra \geq
\e^{\mp\beta\la W\ra}$. From this follows the fundamental inequality
\begin{align} \label{meanworkineq}
 \la W\ra_1 \le \df \le \la W\ra_0,
\end{align}
which shows that the values of the average work in forward and reverse
direction constitute an upper and a lower bound on $\df$,
respectively.

Concerning one-sided estimates of $\df$, the targeted forward and
reverse estimators \gl{dfhat0targ} and \gl{dfhat1targ} can be written
$\dfhatf=-\frac{1}{\beta}\widebar{\e^{-\beta W^0}}$ and
$\dfhatr=\frac{1}{\beta}\widebar{\e^{\beta W^1}}$, where the overbar
denotes a sample average according to a sample $\{W_k^0\}$ and
$\{W_k^1\}$ of forward and reverse work values, respectively.
Similarly to \gl{meanworkineq} one finds the inequalities $\dfhatf \le
\widebar{W^0}$ and $\dfhatr \geq \widebar{W^1}$.

Taking the ensemble averages $\la \dfhat_i \ra_i = \mp\frac{1}{\beta}
\la \ln \widebar{\e^{\mp\beta W^i}} \ra_i$ , $i=0,1$, of the one-sided
estimates and applying Jensen's inequality to the averages of the
logarithms, $\la \ln \widebar{\e^{\mp\beta W^i}} \ra_i \le \ln \la
\widebar{\e^{\mp\beta W^i}}\ra_i=\mp \beta\df$, one obtains
\begin{align} \label{dfbounds}
 \la W\ra_1 \le \la \dfhatr \ra_1 \le \df \le \la \dfhatf \ra_0
 \le\la W\ra_0.
\end{align}
In other words, the forward and reverse estimators are biased in
opposite directions for any finite size $N$ of the work samples, but
their mean values form closer upper and lower bounds on $\df$ than the
values of the mean work do.

So far, we were concerned with one-sided estimates of $\df$, only.
However, the full power of the fluctuation theorem \gl{crooks} will
develop when dealing with a two-sided targeted free energy estimator
where a sample of forward {\em and} reverse work values is used
simultaneously, since the fluctuation theorem relates the forward and
reverse work probability densities to each other in dependence of the
free energy difference.

In the next section, we will drop to mention the target map $\M$
explicitly in order to simplify the notation. For instance, we will
write $p(W|i)$, but mean $p(W|i;\M)$, instead.

\section{Two-sided targeted free-energy estimator}
\label{sec-est}

An important feature of the fluctuation theorem \gl{crooks} is that it
provides a way to answer the following question: Given a sample of
$n_0$ work values $\{W^0_i\}=\{W^0_1,\ldots, W^0_{n_0}\}$ in the
forward direction and a second sample of $n_1$ work values
$\{W^1_j\}=\{W^1_1,\ldots, W^1_{n_1}\}$ in the reverse direction, what
would be the best estimator of $\df$ that utilizes the entire two
samples?

If drawn from an ensemble that consists of forward and reverse work
values, the elements are given by a pair of values $(W,Y)$ of work and
direction, where $Y=0$ indicates the forward and $Y=1$ the reverse
direction. The probability density of the pairs $(W,Y)$ is $p(W,Y)$.
The probability density for the work is $p(W):=p(W,0)+p(W,1)$, and
that for the direction is $p_Y:=\int p(W,Y) dW$.

Bayes theorem,
\begin{align}
 \label{bayes} p(W|Y) p_Y = p(Y|W) p(W),
\end{align}
implies the ``balance'' equation
\begin{align}
 \label{balance} p_1\!\!\int\!\! p(0|W) p(W|1) dW  = p_0\!\!\int\!\! p(1|W) p(W|0) dW .
\end{align}
From the fluctuation theorem \gl{crooks} and Bayes theorem \gl{bayes}
follows
\begin{align} \label{ratioprospective}
 \frac{p(0|W)}{p(1|W)} = \e^{\beta(W-C)}
\end{align}
with
\begin{align} \label{defC}
 C=\df + \frac{1}{\beta} \ln \frac{p_1}{p_0}.
\end{align}
Together with the normalization $p(0|W) + p(1|W) = 1$,
Eq.~\gl{ratioprospective} determines the explicit form of the
conditional direction probabilities \cite{Shirts2003},
\begin{align} \label{probprodef}
 p(Y|W) = \frac{\e^{Y\beta(C-W)}}{1+\e^{\beta(C-W)}}, \quad Y=0,1.
\end{align}
Replacing both, the ensemble averages by sample averages and the ratio
$\frac{p_1}{p_0}$ by $\frac{n_1}{n_0}$, the balance equation, $p_1\la
p(0|W) \ra_1  = p_0\la p(1|W) \ra_0 $, results in the two-sided
targeted free energy estimator, $n_1\widebar{p(0|W^1)}  =
n_0\widebar{p(1|W^0)} $, which reads
\begin{multline} \label{sceq}
 \sum_{j=1}^{n_1} \frac{1}{1+\e^{\beta(\dfhatb+\frac{1}{\beta}
 \ln \frac{n_1}{n_0}-W^1_j)}} \\
 = \sum_{i=1}^{n_0} \frac{1}{1+\e^{-\beta(\dfhatb+\frac{1}{\beta}
 \ln \frac{n_1}{n_0}-W^0_i)}}.
\end{multline}
It is worth to emphasize that this estimator is {\em the} optimal
two-sided estimator, a result that is shown with a constraint maximum
likelihood approach in Appendix \ref{app.likeli}. A derivation of
this estimator is also given by Shirts et al.~\cite{Shirts2003} in the
framework of a maximum likelihood approach.

If samples of $n_0$ forward and $n_1$ reverse work values $\{W_i^0\}$
and $\{W_j^1\}$ are given, but no further information is present, it
is the two-sided estimator \gl{sceq} that yields the best estimate of
the free energy difference with respect to the mean square error. If
needed, the samples $\{W_i^0\}$ and $\{W_j^1\}$ can be obtained
indirectly by drawing samples $\{\x_i\}$ and $\{\y_j\}$ of $\rho_0$
and $\rho_1$ and setting $W^0_i=\dhtilde(\x_i)$ and
$W^1_j=\dhtilde(\phir(\y_j))$.

Opposed to the one-sided estimators \gl{dfhat0targ} and
\gl{dfhat1targ}, the two-sided targeted free energy estimator
\gl{sceq} is an implicit equation that needs to be solved for
$\dfhatb$. Note however that the solution $\dfhatb$ is unique.

Let us mention a subtlety concerning the choice of the ratio
$\frac{p_1}{p_0}$. The mixed ensemble $\{(W,Y)\}$ is specified by the
mixing ration $\frac{p_1}{p_0}$, and by the conditional work
probability densities $p(W|Y)$. With the mixed ensemble we are free to
choose the mixing ratio. For instance, replacing the
ensemble averages in the balance equation \gl{balance} by sample
averages results in an estimator $p_1\widebar{p(0|W^1)}  =
p_0\widebar{p(1|W^0)} $ for $\df$ that depends on the value of
the mixing ratio. This raises the question of the optimal choice for 
$\frac{p_1}{p_0}$. As shown in the Appendix \ref{app.likeli},  it is
optimal to choose the mixing ratio equal to the sample ratio,
$\frac{p_1}{p_0}=\frac{n_1}{n_0}$. A result that may be clear
intuitively, since then the mixed ensemble reflects the actual samples
best.

Other free energy estimators follow, if the
explicit expressions \gl{probprodef} and the definition of the
constant $C$, Eq.~\gl{defC}, are inserted in the balance equation
\gl{balance}. The latter can then be expressed as
\begin{align} \label{bennett}
 \e^{\beta \df} = \e^{\beta C} \frac{\int \frac{1}{1+\e^{\beta(C-W)}}
 p(W|1) dW } {\int \frac{1}{1+\e^{-\beta(C-W)}} p(W|0) dW },
\end{align}
and results in the estimator
\begin{align} \label{dfhatBdef}
 \dfhatB(C) = C + \frac{1}{\beta} \ln \frac{ \frac{1}{n_1}
 \sum_{j=1}^{n_1} \frac{1}{1+\e^{\beta(C-W^1_j)}} }{ \frac{1}{n_0}
 \sum_{i=1}^{n_0} \frac{1}{1+\e^{-\beta(C-W^0_i)}} }.
\end{align}
The non-targeted version of this estimator, i.e. for $\M=\id$, is due
to Bennett \cite{Bennett1976} who used a variational principle in
order to find the estimator for the free energy difference that
minimizes the mean square error.

Equation~\gl{bennett} is an identity for any value of $C$, since with
the ratio $\frac{p_1}{p_0}$ the value of
$C=\df+\frac{1}{\beta}\ln\frac{p_1}{p_0}$ can be chosen arbitrarily.
However, concerning the estimator \gl{dfhatBdef}, different values of
$C$ yield different estimates. Bennett's choice is
\begin{align} \label{cb}
 C_{\text{B}}=\df + \frac{1}{\beta} \ln \frac{n_1}{n_0},
\end{align}
i.e. $\frac{p_1}{p_0} = \frac{n_1}{n_0}$, which results from
minimizing the mean square error $\la (\dfhatB-\df)^2 \ra$, where the
angular brackets denote an average over infinitely many repetitions of
the estimation process \gl{dfhatBdef} with $n_0$ and $n_1$ being
fixed. According to the Appendix \ref{app.likeli}, Bennett's choice is
also optimal for any target map $\M$.

With $C=C_{\text{B}}$, Eq.~\gl{dfhatBdef} has to be solved in a
self-consistent manner which is tantamount to solve the two-sided
targeted estimator \gl{sceq}. In other words, $\dfhatB(C_{\text{B}})$
is the unique root $\dfhatb$ of Eq.~\gl{sceq}.

\section{Overlap measures and mean square errors}
\label{sec-rms}

\begin{figure}
 \includegraphics{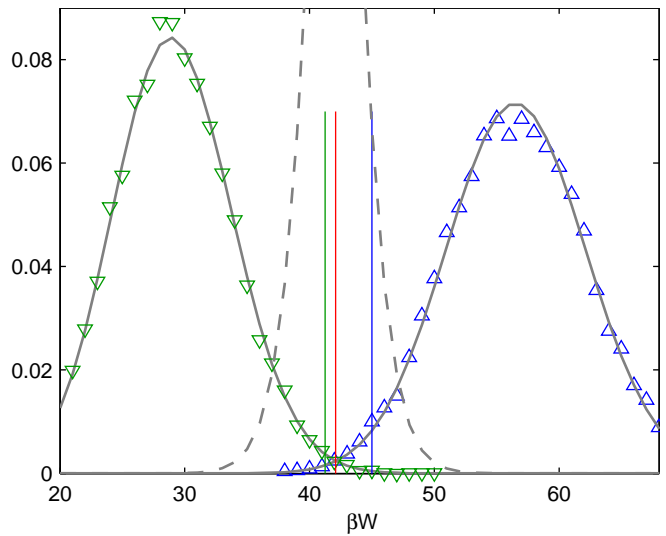}
 \caption{\label{fig:1} Targeted work probability
 distributions for the expansion/contraction of a cavity in an ideal gas
 and the associated overlap distribution. 
 The up (down) triangles display the normalized histogram of a 
 sample of forward (reverse) work values.
 The smooth solid curves are the exact analytic work distributions
 $p(W|0;\M)$ (right) and $p(W|1;\M)$ (left), and the dashed curve
 shows their overlap distribution $p_{\text{ol}}(W|\M)$.
 The straight vertical lines show the
 values of the targeted estimates of $\df$ on the abscissa.
 From left to right: the reverse, the two-sided
 (which is indistinguishable from the exact analytic
 value) and the forward estimate.}
\end{figure}

In this section we introduce measures for the overlap of $\rhotildef$
with $\rho_1$, or, equivalently, of $p(W|0;\M)$ with $p(W|1;\M)$ and
relate them to the mean square error of one- and two-sided
estimators.

The estimators \gl{dfhat0targ}, \gl{dfhat1targ} and \gl{sceq} are
subject to both, bias and variance. Taking both errors into account
results in the mean square error. Let us consider the mean square
errors of the one-sided targeted estimators first. They read $\MSE_0
:= \la (\dfhatf-\df)^2\ra_0 = \la \big( \ln
\widebar{\e^{-\beta(W^0-\df)}} \big)^2 \ra_0$ in forward direction,
and analogously in backward direction. In forward direction, it can be
quantified by expanding the logarithm into a power series about the
mean value of its argument, $\la \widebar{\e^{-\beta (W^0-\df)}} \ra_0
=1$, and neglecting terms of higher order in $\frac{1}{N}$, which
gives
\begin{align} \label{mse0}
 \beta^2\MSE_0 \approx \frac{1}{N} \la \left(\e^{-\beta
 (W-\df)}-1\right)^2 \ra_0.
\end{align}
Equation~\gl{mse0} is valid for a sufficiently large sample size $N$
(large $N$ limit) \cite{Gore2003}. With the use of the fluctuation
theorem \gl{crooks}, the variance appearing on the right hand side of
Eq.~\gl{mse0} can be written $\la \left(\e^{-\beta (W-\df)}-1\right)^2
\ra_0 = \la \e^{-\beta (W-\df)} \ra_1-1 \geq \e^{-\beta (\la W
 \ra_1-\df)}-1$. This yields the inequality
\begin{align} \label{mse0ineq}
 \beta^2\MSE_0 \geq \frac{1}{N}\left( \e^{\beta(\df - \la W\ra_1)}-1
 \right).
\end{align}
In the same manner as above the inequality
\begin{align} \label{mse1ineq}
 \beta^2\MSE_1 \geq \frac{1}{N}\left( \e^{\beta( \la W\ra_0-\df)}-1
 \right)
\end{align}
is obtained for the mean square error $\MSE_1$ of the reverse
estimator $\dfhatr$.

The inequalities \gl{mse0ineq} and \gl{mse1ineq} specify the minimum
sample size $N$ that is required to obtain a forward and reverse
estimate $\dfhat$, respectively, whose root mean square error
$\sqrt{\MSE}$ is not larger than $kT$. Namely, $N\geq\e^{\beta(\df-\la
 W\ra_1)}$ is required for a forward, and $N\geq\e^{\beta(\la
 W\ra_0-\df)}$ for a reverse estimate. Similar expressions are found
in Ref. \cite{Jarzynski2006}. Since the required sample size $N$
depends exponentially on the dissipation, it is good to choose a
target map $\M$ which reduces the dissipation in the opposite
direction.

The dissipation is related to the overlap of $\rhotildef$ with
$\rho_1$. The overlap of two probability densities $\pi_a(z)$ and
$\pi_b(z)$ of a random variable $z$ can be quantified with the
Kullback-Leibler divergence
\begin{align} \label{kld}
 D(\pi_a||\pi_b):=\int \pi_a(z) \ln \frac{\pi_a(z)}{\pi_b(z)}dz,
\end{align}
a positive semidefinite measure that yields zero if and only if
$\pi_a$ is identical to $\pi_b$. Applied to the densities $\rho_1$
and $\rhotildef$, the Kullback-Leibler divergence turns out to be
identical with the Kullback-Leibler divergence of $p(W|1;\M)$ with
$p(W|0;\M)$ and results in the generalized dissipated work in reverse
direction,
\begin{multline} \label{kld1}
 D(\rho_1||\rhotildef) = D(\,p(W|1;\M)\,||\,p(W|0;\M)) \\
 = \beta (\df - \la W \ra_1),
\end{multline}
which is established with the use of Eqs.~\gl{etadrho} and \gl{prdef},
and the fluctuation theorem \gl{crooks}. Similarly, we have
\begin{multline} \label{kld0}
 D(\rhotildef||\rho_1) = D(\,p(W|0;\M)\,||\,p(W|1;\M)) \\
 = \beta (\la W \ra_0 - \df).
\end{multline}
For the one-sided targeted free energy estimators this means that
choosing a target map which reduces the dissipation in the opposite
direction is the same as choosing a target map which enhances the
overlap of $\rhotildef$ with $\rho_1$.

Now, we proceed with the overlap measure and the mean square error of
the two-sided free energy estimator \gl{sceq}. In order to keep the
notation simple, we assume that the samples of forward and reverse
work values are of equal size, $n_0=n_1=N$. (A generalization to
$n_0\not=n_1$ is straightforward possible, but not given in this
paper.)

Consider the overlap density $p_{\text{ol}}(W|\M)$,
\begin{align} \label{oldist}
 p_{\text{ol}}(W|\M) := \frac{1}{\aol} \frac{ p(W|0;\M)
 p(W|1;\M)}{ p(W|0;\M) + p(W|1;\M)},
\end{align}
where the normalization constant $\aol$ reads
\begin{multline} \label{overlapdef}
 \aol=\int \frac{ p(W|0;\M) p(W|1;\M)}{ p(W|0;\M) + p(W|1;\M)} dW \\
 = \int \frac{ \rhotildef(\y) \rho_1(\y)}{ \rhotildef(\y) +
 \rho_1(\y)} d\y.
\end{multline}
$\aol$ is a measure for the overlap area of the distributions and
takes its maximum value $\frac{1}{2}$ in case of coincidence. Using
the fluctuation theorem \gl{crooks}, the two-sided overlap measure can
be written
\begin{align} \label{overlap}
 \aol = \la \frac{1}{1+\e^{\beta(\df-W)}} \ra_1
 = \la \frac{1}{1+\e^{-\beta(\df-W)}} \ra_0.
\end{align}
Comparing Eq.~\gl{overlap} with the two-sided targeted free energy
estimator \gl{sceq}, one sees that the two-sided targeted free energy
estimation method readily estimates the two-sided overlap measure.
The accuracy of the estimate depends on how good the sampled work
values reach into the main part of the overlap distribution
$p_{\text{ol}}(W|\M)$. By construction, the overlap region is sampled
far earlier than the further distant tail that lies in the peak of the
other distribution, cf. Fig.~\ref{fig:1}. This is the reason why the
two-sided estimator is superior if compared to the one-sided
estimators.

In the large $N$ limit the mean square error $\MSE_{\text{B}}(N)=\la
\left( \dfhatb-\df \right)^2\ra$ of the two-sided estimator can be
expressed in terms of the overlap measure and reads
\begin{align} \label{mseb}
 \MSE_{\text{B}}(N) = \frac{1}{N}\left(\frac{1}{\aol}-2\right),
\end{align}
cf. \cite{Bennett1976,Shirts2003}. 
Note that if an estimated value $\aolhat$ is plugged in, this formula
is valid in the limit of large $N$ only, but it is not clear a priori
when this limit is reached. Therefore, we develop a simple
convergence criterion for the two-sided estimate.

\section{Convergence}
\label{sec-con}

In this section, a measure for the convergence of
two-sided estimate is developed, again for the special case $n_0=n_1=N$.
First, we define the estimate $\aolhat$ of the overlap measure $\aol$ with
\begin{align} \label{aolhat}
 \aolhat(N) = \frac{1}{N}\sum_{j=1}^{N}
 \frac{1}{1+\e^{\beta(\dfhatb-W^1_j)}},
\end{align}
which is equal to
$\frac{1}{N}\sum_{i=1}^{N}\frac{1}{1+\e^{-\beta(\dfhatb-W^0_i)}}$,
as we understand the estimate $\dfhatb$ to be obtained according to \gl{sceq} 
with the same samples of forward and reverse work values. 
Since the accuracy of the estimated value $\aolhat$ is unknown, 
we need an additional quantity to compare with.

Another expression for the overlap measure is
\begin{multline} \label{overlapalt}
 \aol = \la \left(\frac{1}{1+\e^{\beta(\df-W)}}\right)^2 \ra_1 \\
 + \la \left(\frac{1}{1+\e^{-\beta(\df-W)}} \right)^2 \ra_0,
\end{multline}
which can be verified with the fluctuation theorem \gl{crooks}. Based
on Eq.~\gl{overlapalt}, we define the overlap estimator of second
order
\begin{multline} \label{aolhatv}
 \aolhatv(N) = \frac{1}{N}\sum_{j=1}^{N}
 \left(\frac{1}{1+\e^{\beta(\dfhatb-W^1_j)}}\right)^2 \\
 + \frac{1}{N}\sum_{i=1}^{N}
 \left(\frac{1}{1+\e^{-\beta(\dfhatb-W^0_i)}}\right)^2.
\end{multline}

Because $\dfhatb$ converges to $\df$, both, $\aolhat$ and $\aolhatv$
converge to $\aol$ in the limit $N\to\infty$. However, the second
order estimator $\aolhatv$ converges slower and is for small $N$
typically much smaller than $\aolhat$, since the main contributions to
the averages appearing in Eq.~\gl{overlapalt} result from work values
that lie somewhat further in the tails of the work distributions.

We use the relative difference
\begin{align} \label{cdef}
 a(N) = \frac{\aolhat-\aolhatv}{\aolhat}
\end{align}
to quantify the convergence of the two-sided estimate $\dfhatb$, where
$\aolhat$, $\aolhatv$ and $\dfhatb$ are understood to be calculated with
the same two samples of forward and reverse work values. 

From Eqs.~\gl{aolhatv}, \gl{aolhat}, and \gl{sceq} follows that $0 \le
\aolhatv \le 2\aolhat$ holds. Hence, the convergence measure $a(N)$
is bounded by
\begin{align}
 -1 \le a(N) \le 1
\end{align}
for any $N$. A necessary convergence condition is $a(N)\to0$. This
means that only if $a(N)$ is close to zero, the two-sided overlap
estimators can have converged. Typically, $a(N)$ being close to zero
is also a sufficient convergence condition. Hence, if $a(N)$ is close
to zero, the mean square error of $\dfhatb$ is given by Eq.~\gl{mseb}
with $\aol\approx\aolhat$. As can be seen from Eq.~\gl{mseb}, the
mean square error and in turn the variance and the bias are reduced
by both, by taking a larger sample size $N$ and by choosing a map $\M$
that enhances the overlap of $\rhotildef$ with $\rho_1$.

With the targeted free energy estimators at hand, together with their
mean square errors, we are now ready to compute free energy
differences numerically.

\section{Numerical examples}
\label{sec-num}

We investigate two numerical applications.
One is the free energy difference of a
fluid subject to the expansion of a cavity which allows the 
comparison with published results \cite{Jarzynski2002}.
The other is the chemical potential of a fluid in the high 
density regime.  

Beneath an ideal gas, the fluid is chosen to be a Lennard-Jones fluid
with pairwise interaction
\begin{align} \label{ljpot}
 V(r_{kl})=4\epsilon \left( \left(\frac{\sigma}{r_{kl}}\right)^{12}
 -\left(\frac{\sigma}{r_{kl}}\right)^{6} \right),
\end{align} 
where $r_{kl}$ is the distance between the k-th and l-th particle,
$r_{kl}=|\rr_k-\rr_l|$. The parameters used are those of argon,
$\sigma =3.542$\,\AA\ and $\epsilon/k=93.3$\,K \cite{Reid1977}.

In all applications, the samples from the 
densities $\rho_0$ and $\rho_1$ are simulated with the Metropolis
algorithm \cite{Metropolis1953}.
In order to simulate macroscopic behavior with a small number $N_p$ 
of particles, periodic
boundary conditions and the minimum image convention
\cite{Frenkel2002} are used. Pairwise interactions are truncated at
half of the box length $R_{\text{box}}=L/2$, but are not shifted,
and the appropriate cut-off corrections are applied \cite{Frenkel2002}.

\subsection{Expansion of a cavity in a fluid}
\label{sec-cavity}

\begin{figure}
 \includegraphics{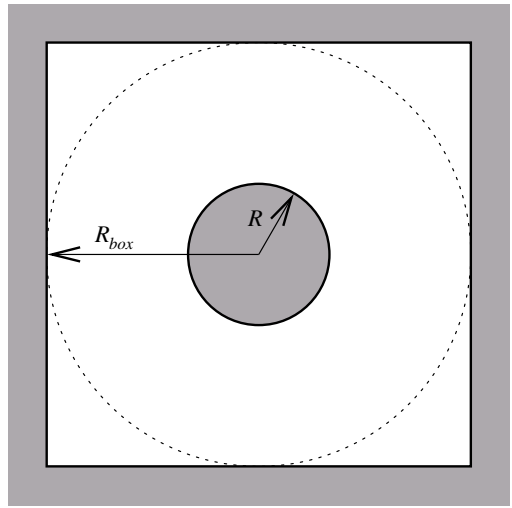}
 \caption{\label{fig:0} The geometric setup.}
\end{figure}

The expansion of a cavity in a fluid is given by the following setup:
Consider a fluid of $N_p$ point molecules with pairwise
interaction $V(r_{kl})$ confined in a cubic box of
side length $2R_{\text{box}}$, but excluded from a sphere of radius $R\le
R_{\text{box}}$, compare with Fig.~\ref{fig:0}. Both,
the box and the sphere are centered at the origin $\rr=0$. A
configurational microstate of the system is given by a set
$\x=(\rr_1,\dots ,\rr_{N_p})$ of particle positions $\rr_k$. Growing
the sphere from $R=R_0$ to $R=R_1$ decreases the volume accessible 
to the particles and the fluid is compressed. We are interested in the increase of
free energy $\df$ subject to the compression of the fluid. Since the
kinetic contribution to the free energy is additive and independent of $R$,
 the difference $\df$ depends only on the configurational
part of the Hamiltonian. The latter reads 
\begin{align}
 H_i(\x) = \begin{cases}
 \sum\limits_{k<l} V(\rr_k,\rr_l) & \text{if}\ \ \x \in \Gamma_i, \\
 \infty & \text{if}\ \ \x \not\in \Gamma_i,
 \end{cases} 
\end{align}
with $i=0,1$. $\Gamma_0$ and $\Gamma_1$ denote the
accessible parts of configuration space of the system $0$ ($R=R_0$)
and $1$ ($R=R_1$), respectively. We assume that $R_0 < R_1$ holds which implies
$\Gamma_1 \subset \Gamma_0$.
 
Drawing a sample $\{\x_k\}$ from $\rho_0$ and applying the traditional
forward estimator \gl{dfhatftrad} results in the following:
 $\e^{-\beta\Delta H(\x_k)}$ takes the values
one and zero depending on whether $\x_k\in \Gamma_1$ or not, i.e.\
whether the region between the two spheres of radius $R_0$ and $R_1$
is found vacant of particles or not.
A comparison with Eq.~\gl{pert0} reveals that $\e^{-\beta\df}$ is
the probability for the spherical shell being observed devoid of
particles \cite{Jarzynski2002}. Hence, the rate of
convergence of $\widebar{\e^{-\beta\Delta H}}$ decreases with the
latter probability and will in general be poor.

Conversely, drawing a sample ${\y_k}$ from $\rho_1$ and applying
the traditional reverse estimator $\dfhatrtrad=\frac{1}{\beta}\widebar{\e^{\beta\Delta H(\y)}}$ (Eq.~\gl{dfhat1targ} with $\phi(\x)=\x$)
figures out to be invalid, because the
term $\e^{\beta\Delta H(\y_k)}$ takes always the value one. In concequence,
the traditional reverse estimator is inconsistent. The deeper reason for this
is that $\Gamma_1\subset\Gamma_0$ holds: Eq.~\gl{rhodrho} is valid only for 
$\x\in\Gamma_1$. By the same reason, the traditional two-sided estimator
is invalid, too. 

The mentioned shortcomings are avoided with a well chosen target
map. Consider mapping each particle separately according to
\begin{align} \label{radmap}
 \phi(\x) = \left(\psi(r_1)\frac{\rr_1}{r_1},\dots,\psi(r_\N)
 \frac{\rr_\N}{r_\N}\right),
\end{align}
where $r_k=|\rr_k|$ is the distance of the $k$-th particle with
respect to the origin, and
$\psi:(R_0,R_{\text{max}}]\to(R_1,R_{\text{max}}]$ is a bijective and
piecewise smooth radial mapping function. In order not to map particles 
out of the confining box, it is required that $\psi(r)=r$ holds for
$r>R_{\text{box}}$. The Jacobian for the radial map \gl{radmap} reads
\begin{align} \label{radjac}
 \Big| \frac{\partial \phi}{\partial \x} \Big| =
 \prod\limits_{j=1}^{N_p} \frac{\psi(r_j)^2}{r_j^2} \frac{\partial
 \psi(r_j)}{\partial r_j}.
\end{align}
(This formula is immediately clear when changing to polar coordinates)
We use the map of Ref.~\cite{Jarzynski2002} which is designed to 
uniformly compress the volume of the shell $R_0< r\le R_{\text{box}}$ 
 to the volume of the shell $R_1<r\le
R_{\text{box}}$. Thus, for $r\in(R_0,R_{\text{box}}]$ the radial mapping function
$\psi(r)$ is defined by
\begin{align} \label{psicavity}
 \psi(r)^3-R_1^3 = c\left( r^3-R_0^3 \right),
\end{align}
with the compression factor
$c=(R_{\text{box}}^3-R_1^3)/(R_{\text{box}}^3-R_0^3)$. 
According to Eq.~\gl{radjac}, we have $\ln K(\x) = \nu(\x) \ln c$, where
$\nu(\x)$ is the number of particles in the shell $R_0 < r \le
R_{\text{box}}$.

\subsubsection{Ideal Gas}

As a first illustrative and exact solvable example we choose the fluid
to be an ideal gas, $V(r_{kl})=0$. In this case the free energy
difference is solely determined by the ratio of the confined volume
$V_i=8R_{\text{box}}^3-\frac{4}{3}\pi R_{i}^3$, $i=0,1$, and is given
by $\beta\df=-N_{p}\ln \left(V_1/V_0\right)$. Using the radial map
\gl{psicavity}, the work in forward direction as a function of $\x$
reads $\dhtilde(\x)=-\frac{1}{\beta} \nu(\x) \ln c$ and takes discrete
values only, as $\nu(\x)=n$ holds with $n \in \left\lbrace
 0,1,\dots,N_p\right\rbrace$. Consequently, the probability
$p(W_n|0;\M)$ of observing the work $W_n=-\frac{n}{\beta} \ln c$ in
forward direction is binomial,
\begin{align} \label{workprobidgas}
 p(W_n|0;\M) = \binom{N_p}{n} q_0^n (1-q_0)^{N-n},
\end{align}
where $q_0 = \frac{4}{3}\pi(R_{\text{box}}^3-R_0^3)/V_0$ is the
probability of any fixed particle to be found in the shell $R_0 < r
\le R_{\text{box}}$. In analogy, the probability distribution
$p(W_n|1;\M)$ for observing the work $W=W_n$ in reverse direction is
given by replacing the index $0$ with $1$ in \gl{workprobidgas}.
Finally, the work probability distributions (rather then the
densities) obey the fluctuation theorem \gl{crooks} for any
$n=0,1,\dots,N_p$,
\begin{align}
 \frac{p(W_n|0;\M)}{p(W_n|1;\M)} = \frac{1}{c^n} \left(
 \frac{V_1}{V_0} \right)^{N_p}=\e^{\beta(W_n-\df)}.
\end{align}

A simple numerical evaluation highlights the convergence properties.
Choosing the parameters to be $2R_{\text{box}}=22.28$\,\AA,
$R_0=7$\,\AA, $R_1=10$\,\AA, and $N_{p}=125$ ($\beta$ arbitrary), the
free energy difference takes the value $\beta\df=42.1064$. Because
$\e^{-\beta \dhtilde(\x)}$ can take only the numbers zero and one, the
probability of observing a configuration $\x$ with non-vanishing
contribution in the traditional forward estimator of $\df$ is
$\e^{-\beta\df} \approx 10^{-19}$. Hence, in practice it is impossible to
use the traditional method successfully, since it would require at
least $N_p\cdot 10^{19}$ Monte Carlo trial moves. However, the targeted
approach already gives reasonable estimates with a sample size of just a few
thousands. Figure \ref{fig:1} shows estimates of the targeted
work probability distributions for samples of size $N=10^4$ 
 from $\rho_0$ and $\rho_1$ each. While the forward distribution
$p(W|0;\M)$ is obviously well sampled in the central region, the
sampling size is too small in order to reach the small values of
$\beta W$ where the reverse distribution $p(W|1;\M)$ is peaked. 
Exactly the latter values would be required for  an {\em
 accurate} exponential average in the targeted forward estimator Eq.~\gl{dfhat0targ}. Therefore,
the targeted forward estimate of $\df$ is still inaccurate; it yields
$\beta\dfhatf=45.0 \pm 0.3$. The same is true for the targeted reverse
estimate \gl{dfhat1targ}  which gives
$\beta\dfhatr=41.3 \pm 0.5$. The errors are calculated using root mean
squares and propagation of uncertainty. A  more accurate estimate follows
from the targeted two-sided
estimator \gl{sceq} which  yields $\beta\dfhatb = 42.1 \pm 0.1$ ($n_0=n_1=N$).
This is clear, as for the two-sided estimate it is sufficient yet that the 
forward and reverse work-values sample the region where the overlap distribution
$p_{\text{ol}}(W|\M)$, Eq.~\gl{oldist}, is peaked, which is obviously the case, 
cf. Fig.~\ref{fig:1}.

\subsubsection{Lennard-Jones fluid}

\begin{figure}
 \includegraphics{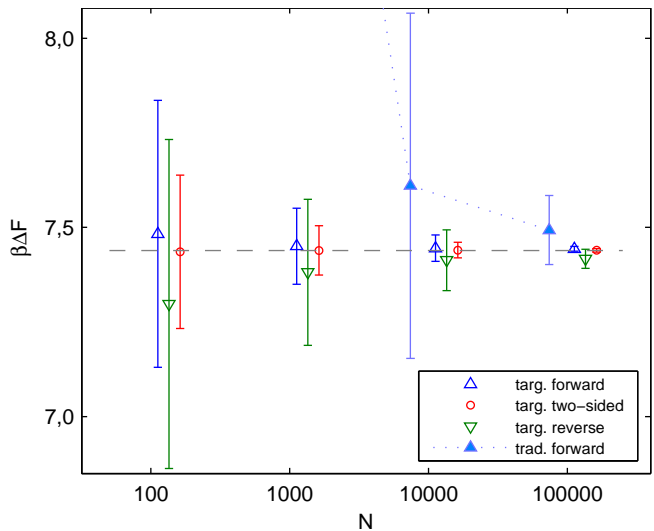}
 \caption{\label{fig:2} Free energy estimates for the expansion 
 of a cavity in a Lennard-Jones fluid. Shown are
 average values of traditional and
 targeted estimates of $\df$ in dependence of the
 sample size $N$, with an errorbar of one standard-deviation. 
 In order to distinguish the data points those
 corresponding to targeted estimates are shifted to the right and
 are spread, whereas those corresponding to traditional estimates
 are shifted to the left. For example, all four data points in the
 vicinity of $N=10000$ refer to $N=10000$. The dashed horizontal
 line represents a targeted two-sided estimate with
 $N=7.5\cdot10^5$, see Table \ref{tab0}.}
\end{figure}

We now focus on particles with Lennard-Jones interaction \gl{ljpot}.
The parameters are chosen to coincide with those of Ref.~\cite{Jarzynski2002}, 
i.e. $2R_{\text{box}}=22.28$\,\AA,
$R_0=9.209$\,\AA, $R_1=9.386$\,\AA, $N_p=125$, and $T=300$\,K. In
Lennard-Jones units, the reduced densities $\rho_i^*=N_p/V_i\cdot\sigma^3$
of the systems $0$ ($R=R_0$) and $1$ ($R=R_1$) are $\rho_0^*=0.713$
and $\rho_1^*=0.731$, respectively, and $T^*=1/(\beta\epsilon)=3.215$
holds for both. If we had an ideal gas, the probability of observing
the space between the spheres of radius $R_0$ and $R_1$ to be vacant
of particles would be $(V_1/V_0)^{N_p}=0.044$. Because of the strong
repulsive part, this probability is much smaller in case of a dense
Lennard-Jones fluid.
 
We generate samples of $\rho_0(\x)$ and $\rho_1(\x)$ with a Metropolis
Monte Carlo simulation. Each run starts with $1000$ equilibration
sweeps, followed by the production run. In the production run the
configurational microstate $\x$ is being
sampled every $4$-th sweep only in order to reduce correlations between
successive samples. The use of decorrelated data is of particular
importance for the self-consistent two-sided estimate $\dfhatb$, because
it depends intrinsically on the ratio $\frac{n_1}{n_0}$ of the numbers of {\em
 uncorrelated} samples, cf. Eq.~\gl{sceq}.

Fig.~\ref{fig:2} gives an overview of independent
runs with different sample sizes $N$, where the one- and 
two-sided targeted estimators can be compared with each
other and with the traditional forward
estimator. Displayed is the estimated mean $\widebar{\widehat{\Delta F}(N)}$ in
dependence of the sample size $N$.
The error bars reflect the estimatet standard deviation $\widebar{\left( \widehat{\Delta F}(N)-\widebar{\widehat{\Delta F}(N)} \right)^2 }^{1/2}$. Each mean and each standard-deviation
is estimated using $z(N)$ independent estimates $\widehat{\Delta F}(N)$.
In ascending order of $N$, $z(N)$ reads $400$, $100$, $20$,
$7$. For the two-sided estimates, $n_0=n_1=N$ is used and
Eq.~\gl{sceq} is solved. 

Note that the theoretical mean of {\em
 traditional} forward estimates of $\df$ is infinite {\em for any
 finite N}, because of the finite probability of observing a sequence of length $N$
of solely vanishing contributions to the exponential average
$\widebar{\e^{-\beta\Delta H}}$. Strictly spoken, the estimator $\beta\dfhatftrad =
-\beta^{-1}\ln \widebar{\e^{-\beta\Delta H}}$ is not well defined,
because $\Gamma_1\subset\Gamma_0$. Nevertheless, in
Fig.~\ref{fig:2} there are two finite {\em observed} mean values of
traditional forward estimates displayed, what by
no means is a contradiction. Infinite values are observed in the cases
where $N<10^4$ holds. This is symbolized by the rising dashed line.
The mentioned ill-definiteness of the traditional estimator is removed
by using the map \gl{psicavity}. Figure \ref{fig:2} shows that all
three targeted estimators are consistent even for small $N$ in the
sense that the error bars overlap. Whereas the targeted forward 
 and reverse estimators show to
be decreasingly biased with increasing $N$, the targeted two-sided
estimator  does not show any noticeable bias at all.
This example demonstrates how worth it can be to take all three estimators,
forward, reverse, and two-sided, into account. The one-sided estimators
are biased in opposite directions and may serve as upper and lower
bounds for $\df$, Eq.~\gl{dfbounds}, whereas the two-sided is
placed in between the one-sided.
 
We conclude this example with explicit estimates obtained from a
single run with $N=750000$, that are summarized in Table \ref{tab0}.
The errors are derived using block averages \cite{Flyvbjerg1989} and
propagation of uncertainty.
\begin{table}[h]
 \caption{Cavity in a Lennard-Jones fluid. Estimated free energy differences 
 $\beta\dfhat$ for the expansion of a cavity, using targeted and traditional
 estimators. $N=7.5\cdot 10^5$.}
 \begin{tabular}{r|l} \label{tab0} 
 Method & $\beta\dfhat$ \\
 \hline	 
 traditional forward & $7.500\pm 0.050$ \\
 targeted forward & $7.442\pm 0.005$ \\
 targeted two-sided & $7.439\pm 0.002$ \\
 targeted reverse & $7.420\pm 0.010$ 
 \end{tabular}
\end{table} 

\subsection{Chemical potential of a homogeneous fluid}
\label{sec-mu}

Consider a fluid of $N_p$ particles confined within a cubic box of
volume $V_c=(2\Rb)^3$ with pairwise interaction $V(r_{ij})$. The
configurational Hamiltonian for the $N_p$-particle system at
$\x=(\rr_1,\dots,\rr_\N)$ reads
\begin{align} \label{hN}
 H_\N(\x)=\sum\limits_{_{i<j}^{i,j}}^{\N} V(r_{ij}).
\end{align}
 The configurational
density for the $N_p$-system is given by
\begin{align} \label{rhoN}
 \rho_\N(\x)=\e^{-\beta H_\N(\x)} / Z_\N,
\end{align}
with the partition function $Z_\N = \int \e^{-\beta
 H_\N(\x)} d\x$. Now consider one particle is added: the position of
this new particle may be $\rr_\Npl$. The equilibrium density of the
($N_p+1$)-particle system reads
\begin{align} \label{rhoNp1}
 \rho_\Npl(\x)=\e^{-\beta H_\Npl(\x,\rr_\Npl)} / Z_\Npl.
\end{align}
Taking the ratio of the densities \gl{rhoN} and \gl{rhoNp1} leads to
Widoms particle insertion method \cite{Widom1963} for estimating the
excess chemical potential $\muex$ of the $N_p$-system, defined as the
excess of the chemical potential $\mu$ to that of an ideal gas at the
same temperature and density. For sufficiently large $N_p$, $\muex$
can be approximated with
\begin{align} \label{mudef}
 \muex = -\frac{1}{\beta} \ln \frac{Z_\Npl}{Z_\N V_c}.
\end{align}
Turning the tables, we use Eq.~\gl{mudef} to be the definition of the
quantity $\muex$. The particle insertion method inserts at a random position
an extra particle to the $N_p$-system and measures the increase of energy
that results from this particle. Since we consider a homogeneous fluid, we may as
well fix the position of insertion arbitrarily, for instance at the origin, what
is done in the following. We define system $1$ through the configuration-space
density $\rho_1(\x)$ at follows:  
\begin{align}
 \rho_1(\x)= V_c \int\delta(\rr_\Npl) \rho_\Npl(\x,\rr_\Npl)
 d\rr_\Npl.
\end{align}
The factor $V_c$ ensures normalization. Written in the usual form
$\rho_1(\x) = \e^{-\beta H_1(\x)}/Z_1$, we have
\begin{align}
 H_1(\x) = H_\N(\x)+\sum\limits_{k=1}^{\N} V(r_k)
\end{align}
and $Z_1 = Z_\Npl/V_c$. System $1$ can be understood as an equilibrium
system of $N_p$ interacting particles in the external potential
$\sum\limits_{k=1}^{\N}V(r_k)$, due to one extra particle fixed at the
origin $\rr=0$. Further, we identify system $0$ with the $N_p$-particle
system and rewrite 
\begin{equation}
\rho_0(\x) = \rho_\N(\x), \quad  H_0(\x)= H_\N(\x)
\end{equation}
and $Z_0 = Z_\N$. The ratio of
$\rho_0$ and $\rho_1$ has the familiar form of Eq.~\gl{rhodrho}, with
$\df$ being identical to $\muex$,
\begin{align}
 \frac{\rho_0(\x)}{\rho_1(\x)}=\e^{\beta (\Delta H(\x)-\muex)}.
\end{align}
The energy difference $\Delta H(\x)=H_1(\x)-H_0(\x)$ is the
increase of energy due to an added particle at the
origin $\rr=0$,
\begin{align}
 \Delta H(\x)= \sum\limits_{k=1}^{\N}V(r_k).
\end{align}
Assume a finite potential $V(r)$  for non-vanishing $r$ 
(i.e. no hard-core potential), but with a
strong repulsive part for $r \to 0$ (a so-called soft-core potential),
e.g. a Lennard-Jones potential. In this case, the configuration spaces
of system $0$ and $1$ conicide, i.e. $\Gamma_0=\Gamma_1$.
Thus a traditional estimate of
$\muex$ is in principle valid in both directions, forward and reverse.
In forward direction we have the equivalent to the particle insertion
method \cite{Widom1963}, 
$\beta\muhatf^{\text{trad}}=-\ln\widebar{\e^{-\beta \Delta H(\x)}}$,
but with fixed position of insertion $\rr=0$.
Here $\x$ is drawn from $\rho_0$ and we will typically find a particle
in a sphere of radius $\bar{r}$ centered at the origin. $\bar{r}$ can roughly
be estimated by the mean next-neighbor distance $(V_c/N_p)^{1/3}$ of an
ideal gas. The dominant contributions to the exponential
average come from realizations $\x$ that resemble typical realizations
of system $1$ \cite{Jarzynski2006}. However, typical realizations $\x$
of the system $1$ do not contain any particle within a sphere of some
radius $r_{hc}$ centered at the origin, because of the extra particle
fixed at the origin
and the strong repulsive part of the interaction. $r_{hc}$ may
be regarded as a temperature-dependent effective hard-core radius of
the interaction $\beta V(r)$. We conclude that the insertion method is
accurate and fast convergent only, if $r_{hc}^3<\!\!<\bar{r}^3$, i.e. for
low densities. Concerning the reverse traditional estimator 
$\beta\muhatr^{\text{trad}}=\ln\widebar{\e^{\beta \Delta H(\y)}}$,
where $\y$ is drawn from $\rho_1$, the same argumentation 
reveals the impossibility of obtaining an accurate estimate in this way.
Effectively, the particles of
system $1$ cannot access the 
vicinity of the origin, no matter how large the sample size will be.
In this sense, $\Gamma_1$ can be substituted with an effective
$\Gamma_1^{\text{eff}}\subset\Gamma_1=\Gamma_0$, implying that the
traditional reverse estimator tends to be inconsistent.

\subsubsection{Constructing a map}

\begin{figure}
 \includegraphics{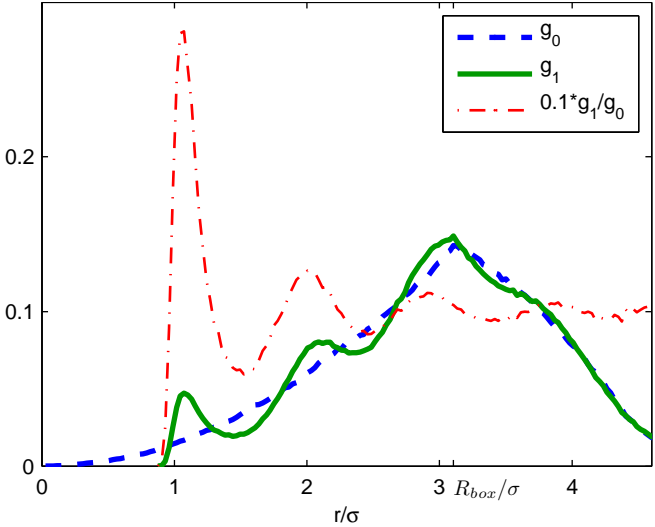}
 \caption{\label{fig:4} The radial densities $g_1(r)$ and
 $g_0(r)$ for a dense Lennard-Jones fluid ($\rho^*=0.9$
 and $T^*=1.2$), estimated from simulated data. The ratio
 $g_1(r)/g_0(r)$ equals the radial
 distribution function $\rdf(r)$.}
\end{figure}

Again, we  use a radial map the particle positions,
$\phi(\x) = (\RR_1,\dots,\RR_\N)$, with
$\RR_k = \psi(r_k) \frac{\rr_k}{r_k}$. In searching a suitable
radial mapping function $\psi(r)$, we are guided by the mean radial
properties of the systems themselves. The radial
probability density $g_0(r)$ of finding a particle in distance $r$
from origin in system $0$ is
\begin{align} \label{g0def}
 g_0(r)= \frac{1}{N_p}\sum\limits_{k=1}^{N_p} \int \delta(r_k-r)
 \rho_0(\x) d\x,
\end{align}
and that for system $1$ is
\begin{align} \label{g1def}
 g_1(r)=\frac{1}{N_p}\sum_{k=1}^{N_p} \int \delta(r_k-r) \rho_1(\x)
 d\x.
\end{align}
Due to the interaction with the extra particle fixed at the origin in
system $1$, $g_1(r)$ will in general be quite different from $g_0(r)$.
The latter is related to a homogeneous fluid and  is
proportional to $~r^2$ (for $r<R_{box}$), whereas the former refers to
an inhomogeneous one and is proportional to $r^2\e^{-\beta V(r)}$ in
the limit $r\to 0$ \cite{Widom1963}. For large $r$, however, the
influence of the extra particle vanishes and $g_1(r) \to g_0(r)$.
Evaluation of the definition \gl{g0def} of $g_0$ yields
\begin{align} \label{g0}
 g_0(r) = \frac{r^2}{V_c} h_0(r),
\end{align}
where $h_0(r)$ accounts for the decay of volume in the corners of the
confining box and is given by $h_0(r)=\iint_{A(r)}\sin\theta d\phi
d\theta $. The integration extends over the fraction of surface $A(r)$
of a sphere with radius $r$ that lies inside the confining box. Note
that $h_0(r)=4\pi$ for $r<\Rb$. In contrast to $g_0$, $g_1$ depends
on the interaction $V(r)$. After some transformations of the right-hand
side of Eq.~\gl{g1def}, $g_1$ can be written
\begin{align} \label{g1}
 g_1(r) = \frac{r^2\e^{-\beta V(r)}}{V_c} h_1(r).
\end{align}
The function $h_1(r)$ can be written (cf. \cite{Widom1963})
\begin{align}
 h_1(r) = \e^{2\beta\muex} h_0(r) \la
 \e^{-\beta\sum\limits_{k=1}^{\scriptscriptstyle N_p-1} \left(
 V(r_k)+V(|\rr_k-\rr_\N|) \right) }\ra_{\scriptscriptstyle
 (N_p-1)},
\end{align}
where the angular brackets denote an average with a $N_p-1$ particle
density according to Eq.~\gl{rhoN} and the vector $\rr_\N$ is
arbitrarily fixed, but of magnitude $r$. Further, the approximation
$V_c^2 Z_{\N-1}/Z_{\Npl} \approx \e^{2\muex}$ is used.

We note that the ratio of $g_1$ with $g_0$ yields the well-known
radial distribution function $\rdf(r)$ of the $N_p+1$-particle fluid, 
\begin{align} \label{g}
 \rdf(r)=\frac{g_1(r)}{g_0(r)}.
\end{align}
Fig.~\ref{fig:4} shows estimates of $g_0$ and $g_1$ for a dense
Lennard-Jones fluid with parameter values of argon (see below
Eq.~\gl{ljpot}), obtained from Monte Carlo simulations.

\begin{figure}
 \includegraphics{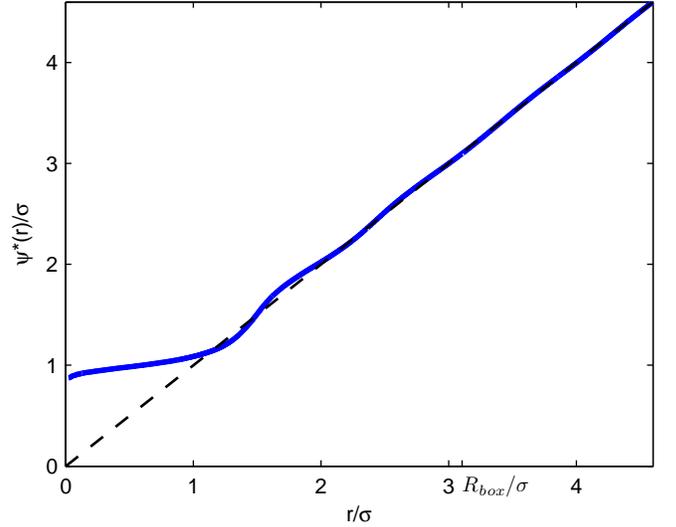}
 \caption{\label{fig:5} Simulated radial mapping function $\psi^*(r)$
 for a dense Lennard-Jones fluid (solid). $\psi^*$ maps the
 radial density $g_0(r)$ to $g_1(r)$, cf. Fig.~\ref{fig:4}. 
 For the ideal gas, $\psi^*$ is the identity map (dashed).}
\end{figure}

Now define a function $\psi^*(r)$ by the requiring that it maps
the mean radial behavior of system $0$ to that of system $1$. This is
done by demanding
\begin{align} \label{psirddef}
 \int\limits_{0}^{\psi^*(r)} g_1(t) dt = \int\limits_{0}^{r} g_0(t)
 dt,
\end{align}
which yields
\begin{align} \label{dpsird}
 \frac{\partial \psi^*}{\partial r} = \frac{g_0(r)}{g_1(\psi^*(r))}.
\end{align}
In the limiting case of an ideal gas, $g_1 = g_0$ holds and the
map becomes an identity, $\psi^*(r)=r$. Of practical interest are
the cases where $g_1$ is unknown and thus Eq.~\gl{psirddef} can not be
used to derive $\psi^*(r)$. However, the function $\psi^*$ can be
estimated with Monte Carlo simulations without knowledge of $g_1$ and
$g_0$ as follows.

Take a sufficiently large amount $n$ of samples
$\x_j=(\rr_{1j},\dots,\rr_{\N j})$, $j=1,\dots,n$, drawn from
$\rho_0(\x)$ together with the same number of samples
$\y_{j}=(\RR_{1j},\dots,\RR_{\N j})$ drawn from $\rho_1(\y)$ 
Calculate the distances to the origin $r_{ij} =|\rr_{ij}|$ and
$R_{ij}=|\RR_{ij}|$ and combine all $r_{ij}$ to the set
$(r_a,r_b,r_c\dots)$, as well as all $R_{ij}$ to the set
$(R_a,R_b,R_c,\dots)$. Provided in both sets the elements are ordered
ascending, $r_a\le r_b \le r_c\le \dots$ and $R_a\le R_b \le R_c \le
\dots$, $\psi^*$ is simulated by constructing a one to one
correspondence $r_a \to R_a$, $r_b \to R_b$, $\dots$ and estimating
$\psi^*(r_{\alpha})$ to be $R_\alpha$, $\alpha=a,b,c,\dots$. In
effect, we have drawn the $r_{\alpha}$ and $R_{\alpha}$ from the
densities $g_0(r)$ and $g_1(r)$, respectively, and have established a 
one-to-one correspondence between the ordered samples. We refer to this
scheme as the simulation of the map of $g_0$ to $g_1$.

The solid curve shown in Fig.~\ref{fig:5} is the result of a
simulation of the function $\psi^*$ for a Lennard-Jones fluid
(parameters of argon, $\rho^*=0.9$, $T^*=1.2$). The corresponding
densities $g_0$ and $g_1$ are plotted in Fig.~\ref{fig:4}.
Noticeable is the sudden "start" of $\psi^*$
with a value of roughly $\sigma$. This is due to the strong repulsive part of
the interaction, that keeps particles in system $1$ approximately a
distance $\sigma$ away from the origin. Therefore, the behavior of
$\psi^*(r)$ for $r\to 0$ is not obtainable from finite-time
simulations. However, the definition of $\psi^*$ implies that for any
soft-core potential $\psi^*(0)=0$ holds. To model $\psi^*$ for small
$r$, the limit $g_1(r)\overset{\scriptscriptstyle r\to 0}{\rightarrow}
a r^2 \e^{-\beta V(r)}4\pi/V_c$ can be used, where $a$ is a constant.
Thus, Eq.~\gl{psirddef} can be written
\begin{align} \label{psirtozero}
 \left[{\psi^*}^{-1}(r)\right]^3 = 3a \int\limits_{0}^{r} {r'}^2
 \e^{-\beta V(r')} dr'
\end{align}
in the limit $r\to 0$, with ${\psi^*}^{-1}$ being the inverse of
$\psi^*$. The constant $a$ is in general unknown, but here it can be
chosen such that is fits continuously to the simulated part of
${\psi^*}^{-1}$.

When the function $\psi^*$ is used in the configuration space map
$\phi$ according to Eq.~\gl{radmap}, then, by definition of $\psi^*$,
the radial density $\tilde{g}_0(r)$ of the mapped distribution
$\rhotildef(\x)$, Eq.~\gl{etadef1}, is identical to the one of
$\rho_1(\x)$:
\begin{align} \label{g0tilde}
 \tilde{g}_0(R)
 &:=\frac{1}{N_p}\sum\limits_{k} \int \delta(|\RR_k|-R)
 \rhotildef(\phi) d\phi \nonumber \\
 &= \int \delta(\psi(r_1)-R) \rho_0(\x) d\x \nonumber \\
 &= \iint \delta(\psi(r)-R)\delta(r_1-r) \rho_0(\x) d\x dr \nonumber \\
 &= \int \delta(\psi(r)-R) g_0(r) dr \nonumber \\
 &= g_1(R).
\end{align}
Therefore we expect that the overlap of the mapped distribution
$\rhotildef$ with $\rho_1$ is larger than the overlap of the
unmapped distribution $\rho_0$ with $\rho_1$. However, it must be
noted that the use of $\psi^*$ in the map $\phi$ is in general valid
only in the limit of an infinite large system ($N,V_c\to \infty$;
$N/V_c=\text{const}$), since we have not yet taken into account the
requirement that particles may not be mapped out of the confining box.
If $\Rb$ is chosen large enough, this might not be a serious problem,
cf. Fig.~\ref{fig:5}.

\subsubsection{Application of the radial map $\psi^*$}

\begin{figure}
 \includegraphics{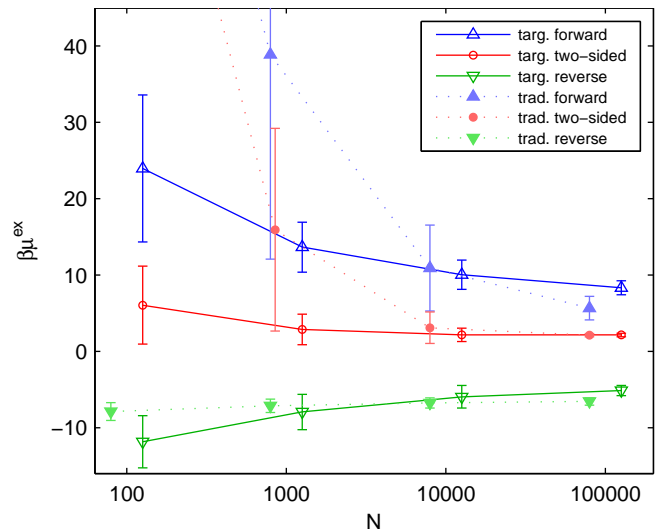}
 \caption{\label{fig:6} Targeted estimates of the excess
 chemical potential $\mu^{ex}$ of a dense Lennard-Jones fluid
 ($\rho^*=0.9$, $T^*=1.2$) compared to traditional estimates. }
\end{figure}

We now apply $\psi^*$ and estimate the
chemical potential of a dense Lennard-Jones fluid ($\rho^*=0.9$, $T^*=1.2$,
parameters of argon) with $\Rb=3.1056 \ \sigma$ and $N_p=216$
particles. Configurations are drawn from $\rho_0$ and $\rho_1$
using a Metropolis algorithm with $7$ decorrelation sweeps between
successive drawings. From every drawn
configuration there results one value for the traditional work and one
for the work related to the map. The usual cut-off corrections
\cite{Frenkel2002} are applied. To avoid mapping particles out of
the confining box, we simulate the map on the interval $0\le r \le R_{box}$
subject to the condition $\psi^*(\Rb)=\Rb$ and use $\psi^*(r)=r$ for $r>\Rb$.
The derivatives of $\psi^*$ and
${\psi^*}^{-1}$ are obtained numerically. For the calculation of the
work values in the simulation, the functions $\psi^*(r)$ and
${\psi^*}^{-1}(r)$ as well as their derivatives are discretized in
steps $\Delta r$ with $\Rb/\Delta r =11\cdot 10^4$.

A comparison of the behavior of the targeted and traditional forward,
reverse and two-sided estimators  in dependence of the sample size $N$ is
given in Fig.~\ref{fig:6} (for the two-sided estimators $n_0=n_1=N$ is
used). Each data point represents the average value
of $z(N)$ independent estimates $\muhat(N)$. The error bars display
one standard deviation. $z(N)$ reads $z(N)=450,250,45, 5$ for $N=100, 1000, 10000, 100000
$, respectively. 

As can be seen from Fig.~\ref{fig:6}, the traditional
one-sided estimators behave quite different. The reverse estimator
converges extremely slow in comparison to the
forward estimator. This can be understood by
comparing the average work values $\widebar{W^i}$ in forward (i=0) and reverse (i=1)
direction, see Table~\ref{tab1}.
\begin{table}[h]
 \caption{Estimatet values of the mean forward and reverse work,
  obtained from $N=10^5$ sampled work values each.
   }
 \begin{tabular}{r|ccc} \label{tab1}
 & $\beta\widebar{W^0}$ & $\beta\widebar{W^1}$  \\
 \hline	 
 traditional & $10^{20}$ & $-9.8$  \\
 targeted & $10^5$ & $-10^6$  \\
 \end{tabular}
\end{table}
Since the absolute value of $\beta\df = \beta\muex$ is small,
the traditional reverse estimator practically never converges, whereas for an
accurate traditional forward estimate we need some $10^5$
work values, cf. Eqs.~\gl{mse0ineq} and \gl{mse1ineq}.
In contrast, the targeted one-sided estimators both show a similar convergence
behaviour if compared with each other. However, the convergence is slow.

The two-sided estimators converge much faster, in particular, the targeted
two-sided estimator converges fastest, see Fig.~\ref{fig:6}.
The convergence of the latter was checked with the convergence measure
$a(N)$, Eq.~\gl{cdef}. A moderate gain in precision for the two-sided
targeted estimator is found if compared to the precision of the two-sided
traditional estimator which can be quantified with the overlap measure
$\aolhat$ \gl{aolhat}. Namely, $\aolhat=1.5\cdot 10^{-4}$ for the 
targeted case, and $\aolhat=1.1\cdot 10^{-4}$ for the traditional case.

We also studied other radial mapping functions
$\psi$. Some of them turned out to give much better results and are
easier to deal with.

\subsubsection{Other radial mapping functions}

\begin{figure}
 \includegraphics{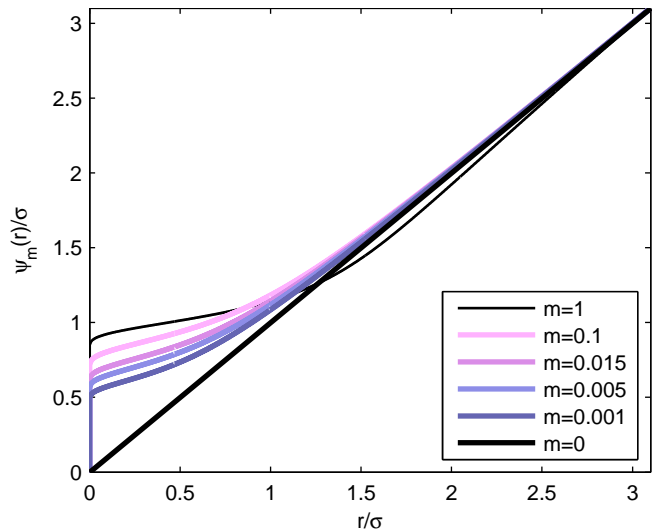}
 \caption{\label{fig:7} Members of the family of radial mapping
 functions $\psi_m$ for the Lennard-Jones potential. For $m\to 0$,
 $\psi_m$ converges to the identity map $\psi_0(r)=r$.}
\end{figure}

The radial mapping function $\psi^*$ was obtained from
simulations, beacause the distribution $g_1(r)$ is analytically unknown.
However, we are free to use any radial mapping function $\psi(r)$ and can thus
in turn fix the function $g_1$ appearing in Eq.~\gl{psirddef}. To do
this, we introduce the normalized, positive definite function
$g'_1(r)$,
\begin{align} \label{g1strich}
 g'_1(r) = \frac{r^2}{c_1} \e^{-\beta(V(r)+Q(r))},\quad r\in [0,R_{box}].
\end{align} 
$Q(r)$ is an arbitrary finite function over $(0,R_{box}]$ and $c_1 = \int_0^{\Rb} r^2\e^{-\beta(V(r)+Q(r))} dr$
a normalization constant. Further, let $g'_0(r)$ be a normalized quadratic density, 
\begin{align}
g'_0(r) = \frac{r^2}{c_0}, \quad r\in [0,R_{box}], 
\end{align}
with $c_0=\Rb^3/3$.

\begin{figure}
 \includegraphics{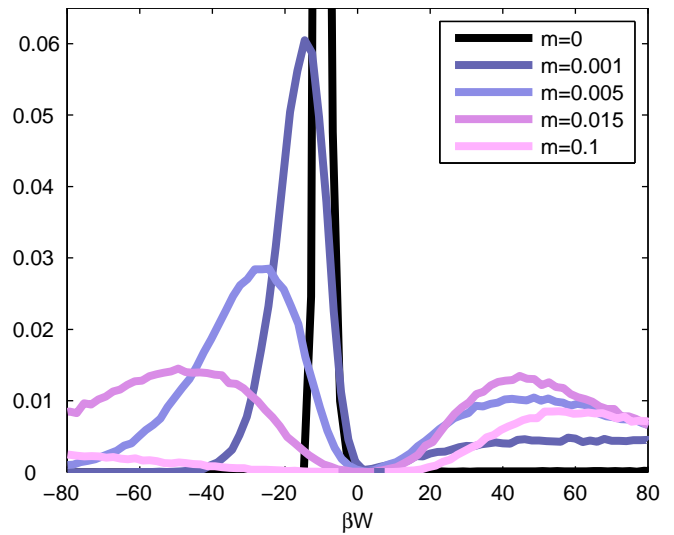}
 \caption{\label{fig:8} Forward (right)
 and reverse (left) work distributions of a Lennard-Jones
 fluid ($\rho^*=0.9$, $T^*=1.2$) for different radial mapping
 functions $\psi_m$. $m=0$ results in 
 the traditional work distributions.}
\end{figure}

The general (monotonically increasing) radial mapping function
$\psi(r)$ can be expressed in terms of the equation
\begin{align} \label{psigendef}
 \int\limits_{0}^{\psi(r)} g'_1(t) dt = \int\limits_{0}^{r} g'_0(t)
 dt
\end{align}
for $r\in[0,\Rb]$. For $r>\Rb$ it shall be understood that $\psi(r)=r$. Given
the function $Q(r)$, $\psi$ and $\psi^{-1}$ are determined uniquely by
Eq.~\gl{psigendef}. An advantage of defining $\psi$ with equation
\gl{psigendef} is that the derivative $\partial\psi/\partial r$ is
given in terms of $V$ and $Q$,
\begin{align}
 \frac{\partial \psi(r)}{\partial r} = \frac{r^2}{\psi(r)^2}
 \e^{\beta \{ V(\psi(r))+Q(\psi(r))-f\} } \ ,
\end{align}
with $f=-\frac{1}{\beta}\ln\frac{c_1}{c_0}$. Using $\psi$ in the
configuration space map $\phi(\x)$ according to Eq.~\gl{radmap} yields
the work function
\begin{multline} \label{workpsigen}
 \dhtilde(\x) = \sum_{i<j}^{(\N)}\left\lbrace
 V(|\RR_i-\RR_j|)-V(|\rr_i-\rr_j|) \right\rbrace \\
 - \sum_{r_i\le\Rb} \left\lbrace Q(\psi(r_i)) -f \right\rbrace.
\end{multline}
Here $\RR_i$ is understood to be $\RR_i=\psi(r_i)\frac{\rr_i}{r_i}$,
and the sum in the second line extends only over those particles for
which $r\le\Rb$ holds. Note that the potential-energy contribution of
the extra particle fixed at the origin is eliminated in the work
function, due to the definition of $\psi$. However, in
Eq.~\gl{workpsigen} we have already assumed $V(r)$ to be cut of at
$r=\Rb$, i.e. $V(r)=0$ for $r \geq \Rb$. Otherwise we had to add
$\sum_{r_i>\Rb}V(\psi(r_i)) = \sum_{r_i>\Rb}V(r_i)$ to
Eq.~\gl{workpsigen}.

\subsubsection{A family of maps}

\begin{figure}
 \includegraphics{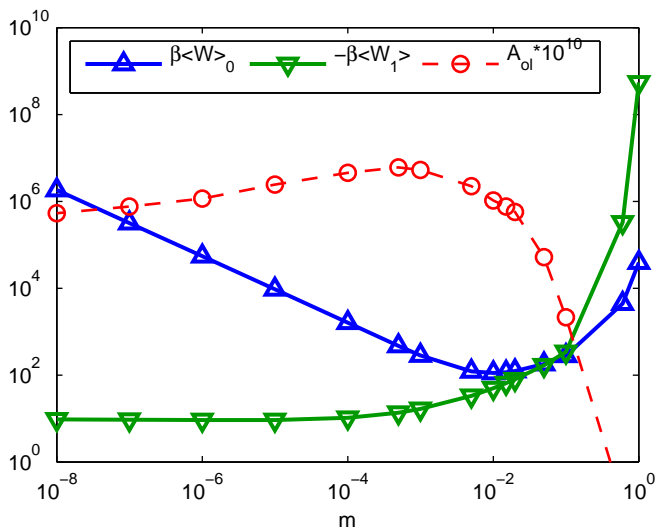}
 \caption{\label{fig:9} The average generalized work $\la W\ra_0$
 and $\la W\ra_1$ in forward  and reverse  direction, respectively, 
 and the two-sided overlap measure $\aol$ in dependence of the mapping 
 parameter $m$. The forward dissipation is reduced up to
 $18$ orders of magnitude if compared with the traditional dissipation,
 cf. Tab.~\ref{tab1}. Among the one-sided estimators the best is 
 found for $m=0$ and in forward direction. The optimal two-sided estimator results from
 using the $m$ that maximizes $\aol$.}
\end{figure}

We now introduce a family $\{\psi_m\}$ of radial mapping functions,
where each member $\psi_m$ is defined by Eq.~\gl{psigendef} with the
choice
\begin{align}
 Q(r)=(m-1)V(r)
\end{align}
in the expression \gl{g1strich}. Useful maps are 
obtained for $m\in[0,1]$. Fig.~\ref{fig:7} depicts some members of
the family $\{\psi_m\}$ for Lennard-Jones interaction (with parameters
of argon). Again, we apply these
functions discretized (in steps $\Delta r$ with $\Rb/\Delta r=11\cdot
10^4$) to the calculation of the targeted forward and reverse work
$\dhtilde(\x)$ and $\dhtilde(\phir(\x))$. Any pair
of forward and reverse targeted work distributions belonging to the
same value of $m$ obeys the fluctuation theorem \gl{crooks}.
In particular they cross at $W=\muex$ ($\df=\muex$ here). Nevertheless,
the shape of these distributions is sensitive to the value of $m$.
This is demonstrated in Fig.~\ref{fig:8}. There, normalized
histograms of $\beta W$ are shown. They result from $10^4$
work values for per $m$ and per direction. We emphasize that all of the
targeted forward (reverse) work values were obtained with {\em one}
sample of $N=10^4$ configurations $\x$ from $\rho_0$
($\rho_1$). 

Instructive is the comparison of the mean work $\la W
\ra$ related to different values of $m$.
In Fig.~\ref{fig:9}  estimated values of mean work are shown in
dependence of $m$. From these values one sees that the dissipation
is minimal for $m=0$ in the reverse direction. Therefore, 
the best one-sided targeted estimate of $\muex$ among
the family $\{\psi_m\}$ is obtained with $m=0$ in forward direction,
i.e. with the traditional particle insertion.
\begin{figure}
 \includegraphics{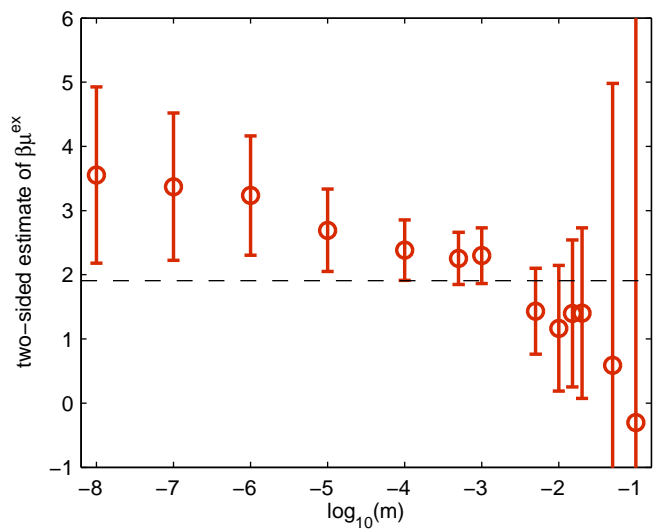}
 \caption{\label{fig:10} Two-sided estimates of $\muex$ as function
 of the mapping parameter $m$ out of $n_0=n_1=N=10^4$ work values
 in both directions for each $m$. The value of the traditional
 estimate ($m=0$) is $\muhatb=4.0\pm2.0$. The error bars show the
 square root of the estimated mean square errors $\MSE_{01}$. For
 comparison, the dashed line represents a two-sided estimate with
 $N=10^6$ and $m=0.0005$ (standard-deviation $0.03$).}
\end{figure}
 \begin{figure}
 \includegraphics{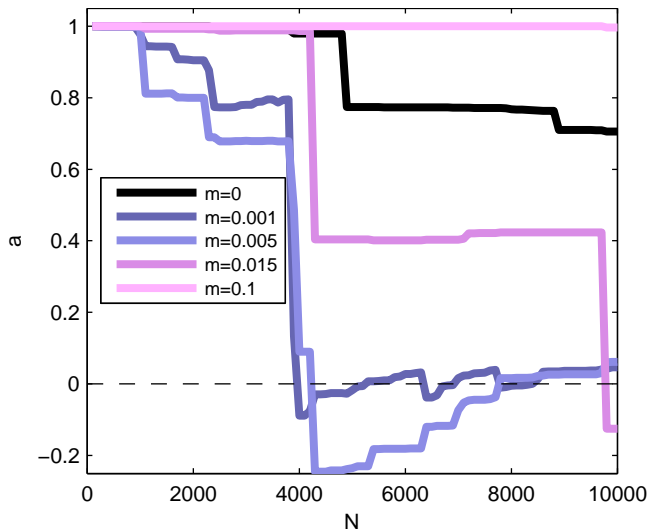}
 \caption{\label{fig:11} Convergence measure $a(N)$ of two-sided
 estimates for some  parameter values $m$, depending
 on the sample size $N$. A faster decrease of $a$ towards the
 value $0$ indicates a faster convergence of
 the two-sided estimator.}
\end{figure}
However, the same is not true for two-sided estimates. 
Using the same data as before and performing
two-sided estimates with $N=10^4$ work values in each direction,
we obtain the displayed values $\muhatb$ of
Fig.~\ref{fig:10}. In order to compare the performance of two-sided 
estimators for different maps, we estimate the overlap measures 
$\aol$. The latter are shown in Fig.~\ref{fig:9}. The maximum
value for $\aol$ is found with $m$ being $0.0005$.
This indicates that $m \approx 0.0005$ is the optimal
choice for $m$. The estimates $\aolhat$ are used to calculate
the mean square errors $\MSE_{01}$ of the
estimates $\muhatb$. The square roots of the $\MSE_{01}$ enter 
in Fig.~\ref{fig:10} as error bars.

We are left to check the convergence properties of two-sided estimators.
Fig.~\ref{fig:11} displays the convergence measure $a(N)$ for some
parameter values $m$. Best convergence is found for $m=0.0005$ (not shown in
Fig.~\ref{fig:11}, but very similar to $m=0.001$). The same value of
the mapping parameter $m$ was found to maximize the overlap $\aol$.

Employing the optimal value $0.0005$ for the mapping-parameter and
using $N=10^6$ forward and reverse samples, we have computed the chemical
potential. The results are given in Table \ref{tab2}.
The listed error is the square root of the $\MSE_{01}$ according to
Eq.~\gl{mseb} with $\aol=\aolhat$. This is justified with the observed
values of the convergence measure $a$ which are listed in the table, too.
\begin{table}[h]
 \caption{Two-sided estimates 
 $\muhatb$ of the excess chemical
 potential of a Lennard-Jones fluid ($\rho^*=0.9$, $T^*=1.2$).
 Also listed is the two-sided overlap measure $\aol$ and the 
 convergence measure $a$. For the
 targeted estimate the radial mapping function $\psi_m$ with $m=0.0005$
 is used. The number of work values in each direction is $N=10^6$ and
 the number of particles in the simulation $N_p=216$.}
 \begin{tabular}{r|ccc} \label{tab2}
 & $\beta\muhatb$ & $10^4\aolhat$ & $a$ \\
 \hline
 traditional & $1.88\pm0.08$ & $1.2$ & $0.05$ \\
 targeted & $1.91\pm0.03$ & $9.5$ & $-0.02$ \\
 \end{tabular}
\end{table}

It should be mentioned that the optimal value of $m$ found here is not
universal, but depends on the density $\rho^*$. If another value is chosen
for $\rho^*$, the optimal $m$ can again be found from numerical simulations.
Note that the maps used here can
be applied to simulations where particles are inserted and deleted at
random \cite{Widom1963}, too. One simply has to use the point of
insertion (deletion) as temporary origin of the coordinate system and
apply the map there. This might enhance the efficiency of the
simulation.

\section{Conclusion}

The central result of this paper,  a fluctuation theorem for generalized work
distributions, allowed us to establish 
an optimal targeted two-sided estimator of the free energy difference $\df$.
We have numerically tested this estimator and found it to be superior
with respect to one-sided and non-targeted estimators.
In addition we have demonstrated that this estimator can be applied successfully
to estimate the chemical potential of a Lennard-Jones fluid in the high
density regime.

In order to use the targeted two-sided estimator it is however crucial
to use a suitable map. We have investigated the construction of maps
and developed appropriate measures which enabled a quantitative comparison of the 
performance of different maps.
Especially, a measure for the convergence of the two-sided estimate was designed.

This paths the way for better results when free energy differences or
chemical potentials need to be estimated numerically.

\section*{Acknowledgments}
We thank Andreas Engel for helpful hints and discussions.

\appendix*

\section{Constraint maximum likelihood derivation of the two-sided
 estimator}
\label{app.likeli}

Deriving the optimal estimator of $\df$, given a collection of $n_0$ forward
$\{W^{0}_i\}$ and $n_1$ reverse $\{W^{1}_j\}$ work-values drawn
from $p(W|0)$ and $p(W|1)$, respectively, leads to Bennett's acceptance ratio
method \cite{Bennett1976} with the target map included.

In Section \ref{sec-est}, the mixed ensemble is introduced,
where the elements are given by pairs of
values $(W,Y)$ of work and direction, and which is specified
by the probabilities of direction $p_Y$ and the densities $p(W|Y)$.
With the mixture ensemble, the mixing ratio $\frac{p_1}{p_0}$ 
can be chosen arbitrarily.
Crucial about the mixture ensemble is that, according to the
fluctuation theorem \gl{crooks}, the analytic form of the conditional
probabilities $p(Y|W)$ can be derived explicitly,
regardless of whether $p(W|Y)$ is known, see Sec.~\ref{sec-est}.
This provides a natural way
to construct a constraint maximum likelihood estimator 
\cite{Aitchison1958,Anderson1972,Prentice1979} for $\df$.

Since it is only possible to draw from the
ensembles $p(W|Y)$, but not from $p(Y|W),\ Y=0,1$, the proper
log-likelihood is
\begin{align} \label{likeli}
 \lnlikeli = \sum_{i=1}^{n_0} \ln p(W^{0}_i|0) + \sum_{j=1}^{n_1} \ln
 p(W^{1}_j|1).
\end{align}
A {\em direct} maximization of \gl{likeli} with respect to $\df$ is
impossible without knowledge of the analytic form of the probability
densities $p(W|Y)$. However, according to Bayes theorem \gl{bayes} 
the log-likelihood can be split into
\begin{align} \label{likeli2}
 \lnlikeli = \lnlikeli_{\text{post}}(\df) + \lnlikeli_{\text{prior}}
 + \lnlikeli_{p_Y}
\end{align}
with
\begin{align} \label{likelipost}
 \lnlikeli_{\text{post}}(\df) = \sum_{i=1}^{n_0} \ln p(0|W^{0}_i) +
 \sum_{j=1}^{n_1} \ln p(1|W^{1}_j),
\end{align}
\begin{align} \label{likeliprior}
 \lnlikeli_{\text{prior}} = \sum_{k=1}^{n_0+n_1}\ln p(W_k),
\end{align}
and
\begin{align} \label{likelifixed}
 \lnlikeli_{p_Y} = n_0 \ln \frac{1}{p_0} + n_1 \ln \frac{1}{p_1},
\end{align}
where the sum in the prior likelihood \gl{likeliprior} runs over all
$n$ observed forward and reverse work values.

Since the definite form of $p(W)$ is unknown, we treat it in the
manner of an unstructured prior distribution and maximize \gl{likeli2}
with respect to the constant $\df$ {\em and} to the function $p(W)$ \cite{Prentice1979}.
Thereby,
\begin{align} \label{conmu}
 1 = \int p(W) dW
\end{align}
and
\begin{align} \label{conlambda}
 p_1 = \int p(1|W) p(W) dW
\end{align}
enter as constraints. Using Lagrange parameters 
$\lambda$ and $\mu$, the constrained log-likelihood reads
\begin{multline} \label{likelic}
 \lnlikeli^{\text{c}} = \lnlikeli + \lambda \big( p_1
 - \int p(1|W) p(W) dW \big) \\
 + \mu \big( 1 - \int p(W) dW \big).
\end{multline}

The conditional direction probabilities $p(Y|W)$ are known explicitly 
in dependence of $\df$, Eq.~\gl{probprodef}, and their partial derivatives 
read $\frac{1}{\beta} \frac{\partial}{\partial \df} \ln p(0|W) = -p(1|W)$
and $\frac{1}{\beta} \frac{\partial}{\partial \df} \ln p(1|W) = p(0|W)
= 1 - p(1|W)$. This allows to extremize the constraint log-likelihood
\gl{likelic} with respect to $\df$,
\begin{multline} \label{vardf}
 0 = \frac{1}{\beta} \frac{\partial}{\partial \df}
 \lnlikeli^{\text{c}} = n_1 - \sum_{k=1}^{n_0+n_1} p(1|W_k) \\
 - \lambda \int \big( 1-p(1|W) \big) p(1|W) p(W) dW.
\end{multline}
Extremizing the conditional likelihood \gl{likelic} with respect to
the function $p(W)$ gives
\begin{align} \label{varpw1}
 \begin{split}
 0 &= \frac{\delta}{\delta p(W)}\lnlikeli^{\text{c}} \\
 &= \frac{1}{p(W)} \sum_{k=1}^{n} \delta(W-W_k) - \lambda p(1|W) - \mu
 \end{split}
\end{align}
which can be solved in $p(W)$,
\begin{align} \label{varpw2}
 p(W) = \frac{\sum_{k} \delta(W-W_k)}{\lambda p(1|W)+\mu},
\end{align}
or written as
\begin{align} \label{varpw3}
 \lambda p(1|W) p(W) = -\mu p(W) + \sum_{k} \delta(W-W_k).
\end{align}

If interested in the values of the Lagrange multipliers $\lambda$ and
$\mu$, one multiplies Eq.~\gl{varpw1} with $p(W)$ and integrates.
This yields
\begin{align}
 0 = n - \lambda p_1 - \mu.
\end{align}
A second independent equation follows from inserting Eq.~\gl{varpw3}
into Eq.~\gl{vardf} which results in
\begin{align}
 0 = n_1 + \mu - \mu p_1 - n,
\end{align}
and the Lagrange multipliers take the values
\begin{align} \label{lagrange}
 \mu = \frac{n_0}{p_0}, \quad \text{and} \quad \lambda
 = \frac{np_0 - n_0}{p_0p_1}.
\end{align}

With the distribution \gl{varpw2} the constraints \gl{conmu} and
\gl{conlambda} read
\begin{align} \label{conmuex}
 1 = \sum_{k} \frac{1}{\lambda p(1|W_k)+\mu}
\end{align}
and
\begin{align} \label{maxlikelieq}
 p_1 = \sum_{k} \frac{p(1|W_k)}{\lambda p(1|W_k)+\mu}
 = \frac{p_1}{n_1} \sum_{k} p^{\text{B}}(1|W_k),
\end{align}
where $p^{\text{B}}(1|W)$ denotes $p(1|W)$ with $C=\df+\frac{1}{\beta}\ln \frac{n_1}{n_0}$.
Whenever the constraint \gl{maxlikelieq} is fulfilled, the constraint
\gl{conmuex} and the variational equations \gl{vardf} and \gl{varpw1}
are automatically satisfied. In consequence, Eq.~\gl{maxlikelieq} defines
the constrained maximum likelihood estimate of $\df$. Note that
the estimator \gl{maxlikelieq} is {\em independent} of the choice of
$\frac{p_1}{p_0}$. Moreover, Eq.~\gl{maxlikelieq} is equivalent to
Eq.~\gl{sceq} regardless of the choice of $\frac{p_1}{p_0}$.

An alternative derivation of the estimator \gl{maxlikelieq} was
presented by Shirts et al.~\cite{Shirts2003}. There, the specific choice
$\frac{p_1}{p_0} = \frac{n_1}{n_0}$ was {\em necessary}. With this choice,
the Lagrange parameter $\lambda$ is identical to zero. Hence, there is
no need to take any constraint into consideration and the posterior
log-likelihood \gl{likelipost} results directly in the estimator of
$\df$.

\end{document}